\documentclass[a4paper,12pt]{article}
\pdfoutput=1 
\usepackage{jheppub}
\usepackage{amssymb, amsmath}
\usepackage{graphicx}
\usepackage{hyperref}

\begin{document}
\begin{flushright}
INR-TH-2017-002
\end{flushright}
\vspace{-1cm}

\title{Exact solutions and critical chaos in dilaton gravity with a
  boundary}
\author[a,b]{Maxim Fitkevich}
\author[a]{Dmitry Levkov}
\author[c,d,e]{Yegor Zenkevich\footnote{On leave of absence from ITEP,
    Moscow 117218, Russia}}
\affiliation[a]{Institute for Nuclear Research of the Russian Academy
  of Sciences, 60th October Anniversary Prospect 7a, Moscow 117312,
  Russia}
\affiliation[b]{Moscow Institute of Physics and Technology,
  Institutskii per. 9, Dolgoprudny 141700, Moscow Region, Russia}
\affiliation[c]{Dipartimento di Fisica, Universit\`a di Milano-Bicocca,
      Piazza della Scienza 3, I-20126 Milano, Italy}
\affiliation[d]{INFN, sezione di Milano-Bicocca, I-20126 Milano, Italy}
\affiliation[e]{National Research Nuclear University MEPhI, Moscow
      115409, Russia}
\emailAdd{fitkevich@phystech.edu}
\emailAdd{levkov@ms2.inr.ac.ru}
\emailAdd{yegor.zenkevich@gmail.com}

\abstract{We consider $(1+1)$-dimensional dilaton gravity with a 
  reflecting dynamical boundary. The boundary cuts off the region of
  strong coupling and makes our model causally similar
  to the spherically-symmetric sector of multidimensional gravity. We 
  demonstrate that this model is exactly solvable at the classical
  level and possesses an on-shell ${\rm SL}(2,\, \mathbb{R})$
  symmetry. After introducing general classical solution of the model,
  we study a large subset of soliton solutions. The latter
  describe reflection  of matter waves off the boundary at low
  energies and formation of
  black holes at energies above critical. They can   be
  related to the eigenstates of the auxiliary integrable system, the
  Gaudin spin chain. We argue that  despite being exactly solvable,
  the model in the critical regime, i.e.\ at the verge of black hole
  formation, displays dynamical instabilities specific to chaotic
  systems. We believe that this model will be useful for studying
  black holes and gravitational scattering.
 }

\maketitle

\section{Introduction}
\label{sec:intro}
The models of two-dimensional dilaton gravity were popular for
decades~\cite{Giddings:1994pj,  Strominger:1994tn,
  Grumiller:2002nm}. Some of them describe spherically-symmetric
sectors of multidimensional gravities with 
dilaton fields $\phi$ related to the sizes of the extra
spheres\footnote{In particular, gravitational sector of the
  CGHS model~\cite{Callan:1992rs}  can be obtained
  by spherical reduction of   $D$-dimensional gravity at $D
  \to +\infty$~\cite{Grumiller:2002nm}.}. Some others are exactly solvable
at the semiclassical~\cite{Callan:1992rs, Russo:1992ax} or
quantum~\cite{Grumiller:2002nm} levels which makes them valuable for 
studying black holes and gravitational
scattering~\cite{Giddings:2011xs, tHooft:1996rdg, Dvali:2014ila}.

These models become particularly important in the context of
information paradox~\cite{Hawking:1974sw, Hawking:1976ra} 
confronting an apparent  loss of quantum coherence  during black hole
evaporation with the principles of quantum theory. Since unitarity of
quantum gravity is 
strongly supported by the AdS/CFT correspondence~\cite{Maldacena:1997re,
  Maldacena:2001kr}, modern AMPS argument~\cite{Almheiri:2012rt,
  Almheiri:2013hfa} suggests dramatic 
violation of the equivalence principle (``firewalls'') in the vicinity
of old black hole horizons, see~\cite{Mathur:2009hf,
  Braunstein:2009my} for earlier works. This feature, if exists, may
leave ``echoes'' in  the gravitational wave
signal~\cite{Cardoso:2016rao, Cardoso:2016oxy} to be detected by
LIGO~\cite{Abbott:2016blz, Abedi:2016hgu}, cf.~\cite{Ashton:2016xff,
  Abedi:2017isz}. From the theoretical 
viewpoint, further progress can be achieved by understanding unitary
evolution  of black holes outside of the explicit AdS/CFT
framework. This brings us to the arena of two-dimensional models which
may, in addition, clarify relation of black holes to quantum
chaos~\cite{Shenker:2013pqa, Shenker:2014cwa, Polchinski:2015cea, 
  Hashimoto:2016dfz, Maldacena:2015waa, Turiaci:2016cvo},
cf.~\cite{Akhmedov:2015xwa}.

\begin{figure}[t!]

  \vspace{-6mm}
    
  \begin{center}
    \hspace{-0.16\textwidth}
    \begin{minipage}{0.32\textwidth}
      \includegraphics[width=\textwidth,angle=45]{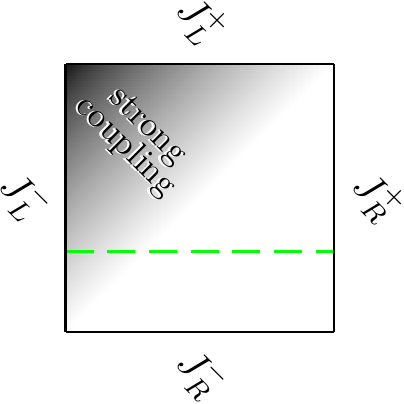}

      \vspace{-5mm}
      \hspace{3.1cm}(a)
    \end{minipage}
    \hspace{0.15\textwidth}
    \begin{minipage}{0.1\textwidth}
      \textcolor{red}{\large $\boldsymbol{\longrightarrow}$}
    \end{minipage}
    \hspace{-0.1\textwidth}
    \begin{minipage}{0.32\textwidth}
      \includegraphics[width=\textwidth,angle=45]{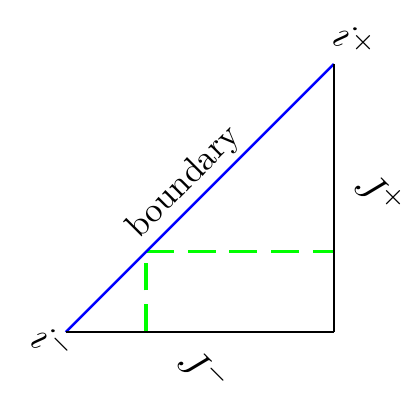}

      \vspace{-5mm}
      \hspace{3.1cm}(b)
    \end{minipage}
  \end{center}
  
  \vspace{-4mm}
  \caption{Penrose diagrams of Minkowski vacuum in the (a) original
    CGHS model and (b) model with a boundary. The dashed lines are
    light rays extending from $J^-$ to~$J^+$.\label{fig:penrose-intro}} 
\end{figure}
Unfortunately, solvable models of two-dimensional dilaton gravity 
essentially 
differ from their multidimensional cousins. Consider  e.g.\ the
celebrated Callan-Giddings-Harvey-Strominger (CGHS)
model~\cite{Callan:1992rs}, see~\cite{Giddings:1994pj,
  Strominger:1994tn} for  reviews. Its two-dimen\-sio\-nal Minkowski 
vacuum in Fig.~\ref{fig:penrose-intro}a, unlike the multidimensional
vacua, has disconnected sets of ``left'' and ``right'' infinities
$J^{\pm}_L$ and  $J^{\pm}_R$, and 
transitions between those are expected~\cite{Ashtekar:2008jd} to be
important for the information loss problem. Besides,
the CGHS model is strongly  coupled~\cite{Das:1994yc} near the
``left'' infinities which puts its  semiclassical results on shaky
ground. It was recently suggested~\cite{Almheiri:2013wka} that due to
the above peculiarities  evaporation of the CGHS black 
holes leads to remnants rather than firewalls.

We consider the modified CGHS model proposed\footnote{Similar models
  appeared recently in the   context of 
  near AdS$_2$ / near CFT$_1$ holography~\cite{Almheiri:2014cka, Jensen:2016pah, 
    Maldacena:2016upp, Engelsoy:2016xyb}.} in~\cite{Chung:1993rf,
  Schoutens:1994st}, see also~\cite{Das:1994yc, Das:1994kr, Strominger:1994xi,
  Bose:1995bk, Bose:1996pi}. The region of strong coupling in this
model is cut off by the reflective dynamical boundary placed at a
fixed value $\phi = \phi_0$ of the dilaton field, see
Fig.~\ref{fig:penrose-intro}b. Parameter $\mathrm{e}^{2\phi_0} \ll
1$ plays the role of a small coupling constant. We explicitly obtain 
reparametrization-invariant action of the model by restricting 
CGHS action to the space-time region $\phi<\phi_0$ and adding
appropriate boundary terms. Note that the original CGHS model is
formally restored  in the limit $\phi_0 \to +\infty$ which shifts the
regulating boundary in Fig.~\ref{fig:penrose-intro}b all the way the
left. We do not consider this limit avoiding potential
problems with strong coupling, cf.~\cite{Russo:1992yh,
  Verlinde:1993sg, Schoutens:1993hu}.  

As an additional bonus, the  above model with a
boundary is causally similar to sphe\-ri\-cal\-ly-\-symmetric
multidimensional gravity, cf.~Fig.~\ref{fig:penrose-intro}b. The price
to pay, however, is nonlinear equation of motion for the boundary which,
if non-integrable, may damage major attractive property of 
the CGHS model~--- its solvability. Note that the previous studies of
this or similar models were relying on
numerical~\cite{Strominger:1994xi, Bose:1995bk, Bose:1996pi, 
  Peleg:1996ce} or shock-wave~\cite{Chung:1993rf, Schoutens:1994st,
  Das:1994kr} solutions.

\begin{figure}[htb]

  \vspace{-15mm}
  
  \centerline{  \hspace{-0.12\textwidth}
    \includegraphics[width=0.36\textwidth,angle=45]{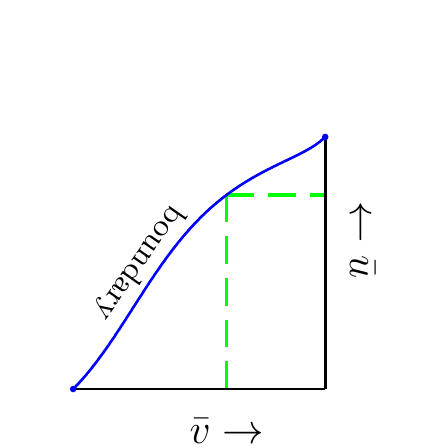}
    \hspace{-0.17\textwidth}
    \includegraphics[width=0.36\textwidth,angle=45]{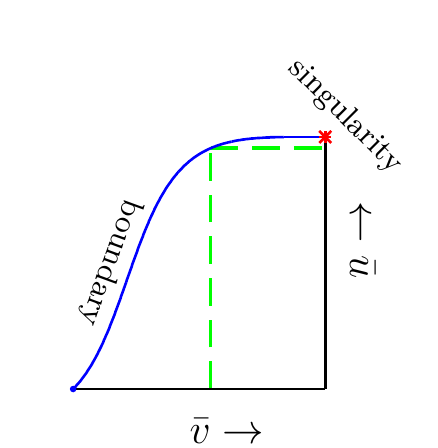}
    \hspace{-0.13\textwidth}
    \includegraphics[width=0.36\textwidth,angle=45]{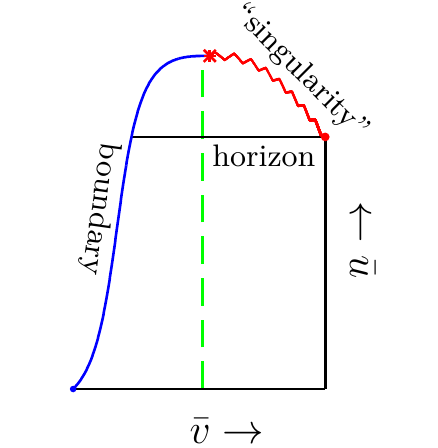}
  }

  \vspace{-5mm}
  \hspace{5mm}(a) low-energy \hspace{25mm} (b) critical
  \hspace{3.4cm}(c) high-energy 

  
  \caption{The simplest exact solution ($s_1 = s_2 =1$) in the
    model with a boundary at different  values of parameters. 
    Finite-range light-cone coordinates $(\bar{u},\,  \bar{v})$
    are used. The centers of the incoming and reflected matter wave
    packets are marked by the dashed
    lines.\label{fig:intro_solutions}}
\end{figure}

In this paper we demonstrate that the CGHS model with a
boundary is exactly solvable at the classical level. We obtain general
solution of the classical field
equations and construct an infinite number of  particular soliton
solutions. The latter describe reflection of matter waves off 
the boundary at low energies and formation of black holes at energies
above some critical values, see  Figs.~\ref{fig:intro_solutions}a and
\ref{fig:intro_solutions}c. Each solution is characterized by $N$ integers or
half-integers $s_1,\, \dots, \, s_{N}$ and the same number of real
parameters. The parameters of the
solitons satisfy inequalities ensuring positivity of energy. 

We establish one-to-one correspondence between the above solitons
and the eigenstates of the auxiliary integrable system~---
the rational Gaudin model~\cite{Gaudin:1976sv, Feigin:1994in, 
  Frenkel:1995zp}. This allows us to classify these solutions and
study their properties. 

We find that equation of motion for the boundary is
invariant under conformal transformations $v\to w(v)$, where $v$ is the
light-cone coordinate, $w(v)$ is an arbitrary function. These
transformations relate physically distinct solutions, and one should not
confuse them, say, with the residual reparametrization symmetry
in~\cite{Das:1994kr, Strominger:1994xi}. In particular,
the transformations from the 
global $\mbox{SL}(2,\mathbb{R})$ subgroup change massless matter
field(s) $f$ of the model as the standard zero-weight fields. They
also  map the solitons into solitons.  The transformations 
with nonzero Schwarzian derivative act  non-linearly on $f$,
and we do not consider them in detail. 

Finally, we study dynamics of the model in the critical regime,
i.e.\ at the verge of black hole formation,
cf.\ Fig.~\ref{fig:intro_solutions}b. We demonstrate that in this
limit scattering of matter waves off the boundary displays
instabilities specific to chaotic systems: the final state of the
process becomes extremely sensitive to the initial Cauchy data. This feature
is in tune with the near-horizon chaos suggested
in~\cite{Polchinski:2015cea}. We argue that it
impedes global integrability of the model, i.e.\ prevents one from 
choosing a complete set of smooth conserved quantities in the
entire phase space. 

In Sec.~\ref{sec:model} we introduce dilaton gravity with a
boundary and study its properties. We construct
exact  solutions in Sec.~\ref{sec:int_sector}. 
Critical chaos is considered in Sec.~\ref{sec:instability}. In
Sec.~\ref{sec:conc} we discuss possible applications of our results.  

\section{The model}\label{sec:model}
\subsection{Adding the boundary}
\label{sec:adding-boundary}
We consider two-dimensional model with classical action
\begin{align}
S=& \int\limits_{\phi < \phi_0}
d^2x\,\sqrt{-g}\left[
  e^{-2\phi}\left(R+4(\nabla\phi)^2+4\lambda^2\right) - (\nabla
  f)^2/2 \right]
\notag\\
&+\int\limits_{\phi = \phi_0}d\tau \, e^{-2\phi}
\left(2K+4\lambda\right)\;,
\label{eq:cghsmirror_action}
\end{align}
where\footnote{We use $(-,\,  +)$ signature and Greek
  indices $\mu,\, \nu,\, \dots = 0,\, 1$. We denote covariant
  derivatives by $\nabla_\mu$ and Ricci scalar by $R$.} the
integrand in the first line is the CGHS 
Lagrangian~\cite{Callan:1992rs} describing interaction of the metric
$g_{\mu\nu}$ 
and dilaton $\phi$ with massless scalar $f$; the dimensionful
parameter $\lambda$ sets the energy scale of the model. In
Eq.~(\ref{eq:cghsmirror_action}) we modified the CGHS action by
restricting integration to the submanifold $\phi<\phi_0$ and adding
the boundary terms\footnote{Similar boundary terms appear in the path
  integral formulation of dilaton gravity~\cite{Grumiller:2007ju}.} at
$\phi = \phi_0$. We introduced the proper time of the boundary $\tau$,
its extrinsic curvature $K=g^{\mu\nu}\nabla_\mu  n_\nu$, and unit
outer normal $n_\mu\,\propto\,\nabla_\mu\phi$.   

In fact, the choice of the boundary action in
Eq.~(\ref{eq:cghsmirror_action}) is limited. First, the
Gibbons-Hawking term with extrinsic curvature ensures consistency of
the gravitational action. Without this term the boundary conditions
following from Eq.~(\ref{eq:cghsmirror_action}) would be incompatible
with the Dirichlet condition $\phi = \phi_0$, see~\cite{Poisson} and
cf.~Appendix~\ref{sec:field-equat-bound}. Second, we assume no direct  
interaction of the matter field $f$ with the boundary. Then the only
natural  generalization of our model would include an arbitrary
constant in the last term of 
Eq.~(\ref{eq:cghsmirror_action}). However, this parameter needs to be
fine-tuned in order to retain Minkowski solution  (see below). Thus,
the action~(\ref{eq:cghsmirror_action}) describing interaction of the
boundary with the gravitational sector of the CGHS
model is fixed~\cite{Chung:1993rf}.

The quantity $\mathrm{e}^{2\phi_0}$ is a coupling constant controlling
loop expansion in the mo\-del~(\ref{eq:cghsmirror_action}). Indeed, change of variables 
$\tilde{\phi} = \phi - \phi_0$, $\tilde{f} = \mathrm{e}^{\phi_0}
f$ brings this parameter in front of the classical action, $S = 
\tilde{S}/\mathrm{e}^{2\phi_0}$. Thus, $\mathrm{e}^{2\phi_0}$ plays
the role of a Planck constant implying that the model is classical
at~$\mathrm{e}^{2\phi_0} \ll 1$.   

It is clear that the bulk equations in the
model~(\ref{eq:cghsmirror_action}) are the same as in the original 
CGHS model~\cite{Callan:1992rs}\cite{Giddings:1994pj, 
  Strominger:1994tn}. However, extremizing the action with respect to
the boundary values of  $g_{\mu\nu}$ and $f$, one also obtains the 
boundary conditions
\begin{equation}
  \label{eq:neumann}
 n^\mu\nabla_\mu \phi=\lambda\;, \qquad n^\mu\nabla_\mu f=0 \qquad
\mbox{at}\qquad \phi=\phi_0\;,
\end{equation}
see Appendix~\ref{sec:field-equat-bound} for details. Note that the 
constant $\lambda$ in the right-hand side of the first equation comes
from the last term in Eq.~(\ref{eq:cghsmirror_action}).  Besides,
the second equation guarantees zero energy flux through the 
boundary. 

Let us now recall~\cite{Callan:1992rs} that linear dilaton vacuum 
\begin{equation}
  \label{eq:10}
  g_{\mu\nu}=\eta_{\mu\nu}\;,\qquad\qquad
  \phi=-\lambda x\;, \qquad\qquad
  f = 0\;,
\end{equation}
satisfies the CGHS equations,
cf.\ Appendix~\ref{sec:field-equat-bound}. In this case the boundary
$\phi = \phi_0$ is static, $x_{\mathrm{boundary}} = -\phi_0/\lambda$,
and the first of Eqs.~(\ref{eq:neumann}) is automatically
satisfied. Note that the Minkowski vacuum~(\ref{eq:10}) is a solution 
in our model due to exact matching between the bulk and boundary terms with
$\lambda$ in the action~(\ref{eq:cghsmirror_action}).

\subsection{Solution in the bulk and reflection laws}
\label{sec:equat-moti-bound} 
The CGHS equations in the bulk are exactly
solvable~\cite{Giddings:1994pj, Strominger:1994tn} in the
light-cone frame $(u,\, v)$, where
\begin{equation}
  \label{eq:6}
  ds^2=-e^{2\rho(u,\, v)}dudv\;.
\end{equation}
Let us review their general solution leaving
technical details to Appendix~\ref{sec:quasi-kruskal-gauge}. In what
follows we fix the remaining gauge freedom in Eq.~(\ref{eq:6}) with
the on-shell ``Kruskal'' condition~$\rho=\phi$. 

In the frame~(\ref{eq:6}) the matter field  satisfies 
$\partial_u \partial_v f = 0$  and therefore splits into a sum of
incoming and outgoing parts,
\begin{equation}
  \label{eq:bulk_scalar}
  f=f_{in}(v)+f_{out}(u)
\end{equation}
The respective energy fluxes are
\begin{equation}
  \label{eq:7}
  T_{vv}(v) = (\partial_v f_{in})^2 
  \qquad\mbox{and}\qquad
  T_{uu}(u)  = (\partial_u f_{out})^2 \;.
\end{equation}
This specifies the Cauchy problem in our model: one prepares $f_{in}$ or
$T_{vv}$ at the past null  infinity and calculates $f_{out}$ or
$T_{uu}$ at $J^+$, see Fig.~\ref{fig:penrose-intro}b.

\begin{figure}[htb]

  
  \vspace{-7mm}
  \centerline{\includegraphics[width=0.32\textwidth,angle=45]{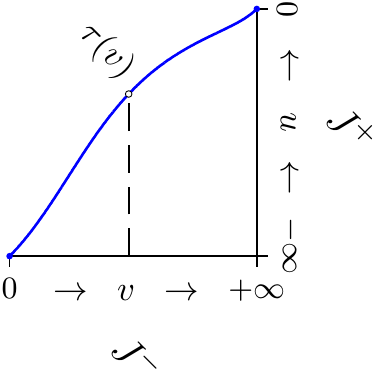}}

  \vspace{-7mm}

  \caption{Penrose diagram showing the ranges of $u$,
    $v$ and definition of~$\tau(v)$.\label{fig:coordinates}} 
\end{figure}

The solution for the scale factor $\rho$ and dilaton field $\phi$ is 
\begin{equation}\label{eq:bulk_gravity}
e^{-2\rho}=e^{-2\phi}=-\lambda^2 vu+g(v)+h(u)\;,
\end{equation}
where
\begin{equation}
  \label{eq:g_h_def}
g(v)=\frac12\int\limits_0^v dv'\int\limits_{v'}^{+\infty}dv'' \,
T_{vv}(v'')\;,\qquad
h(u)=-\frac12\int\limits_{-\infty}^u
du'\int\limits_{-\infty}^{u'}du''\, T_{uu}(u'')\;.
\end{equation}
We fixed the integration constants in these expressions by requiring,
first, that the space-time is flat in the infinite past, i.e.\ no white
hole preexists  the scattering process. Second, we  chose the 
coordinates in such a way that the quadrant $u\in (-\infty;\, 0)$,
$v\in (0;\, +\infty)$ covers all space-time accessible to
the distant observer. In particular, the limits $u\to -\infty$ at
$v>0$ and $v\to  +\infty$ at $u<0$ lead  to $J^-$ and $J^+$,
respectively, see Fig.~\ref{fig:coordinates}.

Now, consider the boundary $\phi = \phi_0$ described by the function ${u
  = U(v)}$ in the ``Kruskal'' coordinates. Substituting the   
bulk solution~(\ref{eq:bulk_scalar}), (\ref{eq:bulk_gravity}) into
the boundary conditions  (\ref{eq:neumann}), one obtains equation for
$U(v)$ and reflection law for the matter field~$f$, 
\begin{equation}
  \label{eq:riccati}
  \frac{dU}{dv} = \frac{e^{2\phi_0}}{\lambda^2}\, \left(\partial_v
    g-\lambda^2 U\right)^2\;, 
  \qquad \qquad
  f_{out}(U(v))=f_{in}(v)\;,
\end{equation}
see Appendix \ref{sec:quasi-kruskal-gauge} for the derivation of these
equations and proof that they are compatible with the definition 
$\phi(U(v),\, v) = \phi_0$ of the boundary. Note that the second of
Eqs.~(\ref{eq:riccati}) relates the incoming and outgoing waves by 
conformal transformation $v\to U(v)$. The first equation implies that
the boundary is always time-like, ${dU/dv>0}$. When rewritten in the
appropriate terms, it coincides\footnote{It does not conform, however,
  with the boundary conditions introduced at one-loop level
  in~\cite{Russo:1992yh, Verlinde:1993sg, Schoutens:1993hu}:
  in the classical model the latter conditions imply that the boundary is 
  space-like.} with the boundary equation obtained
in~\cite{Chung:1993rf, Schoutens:1994st, Das:1994kr} using
energy conservation. 

One easily finds solution in the empty space using
Eqs.~(\ref{eq:riccati}) and (\ref{eq:bulk_gravity})
with $T_{vv} = T_{uu} = 0$,
\begin{equation}
  \label{eq:14}
  U(v) = - \mathrm{e}^{-2\phi_0}/(\lambda^2 v) \;, \qquad 
  \mathrm{e}^{-2\rho} = \mathrm{e}^{-2\phi}=-\lambda^2 uv\;, \qquad f=0\;,
\end{equation}
where the integration constant in the first expression was chosen to
make $U(v)$ smooth and invertible in the interval
$0<v<+\infty$. Solution~(\ref{eq:14}) is the  linear dilaton
vacuum\footnote{Recall that we excluded solutions with eternal black
  holes in Eq.~(\ref{eq:bulk_gravity}).}: coordinate  
transformation
\begin{equation} 
  \label{eq:15}
  \lambda v=e^{\lambda (t+x)}\;, \qquad \qquad
  \lambda u=-e^{-\lambda(t-x)}
\end{equation}
brings it to the standard form~(\ref{eq:10}). In what follows
we impose flat asymptotics (\ref{eq:14}) in the infinite past $v\to
0$, $u\to -\infty$. 

Note that the space-time (\ref{eq:bulk_gravity}) is always flat far
away from the boundary, i.e.\ at large $|u|$ and $v$. Below we 
transform to the asymptotic Minkowski coordinates $(t,\, x)$ using
Eq.~(\ref{eq:15}).  

We have got a receipt for solving the Cauchy
problem in the CGHS model with a boundary. In this case the initial
Cauchy data are represented by the incoming wave $f_{in}(v)$ or its
energy flux $T_{vv}(v)$. One solves Eqs.~(\ref{eq:riccati}) with the
initial  condition (\ref{eq:14}) at $v\to 0$ and finds $U(v)$,
$f_{out}(u)$. The scale factor of the metric, dilaton and matter
fields are then given by Eqs.~(\ref{eq:bulk_gravity})
and~(\ref{eq:bulk_scalar}). 

\subsection{Simple equation for the boundary}
\label{sec:simpl-bound-equat}
One notices that Eq.~(\ref{eq:riccati}) for $U(v)$ is, in fact, a Riccati
equation. The standard substitution 
\begin{equation} 
  \label{eq:13}
  \lambda^2 U = \partial_v g - e^{-2\phi_0}\partial_v \psi/\psi\;,
\end{equation}
brings it to the form of a Schr\"odinger equation for the new
unknown $\psi(v)$,
\begin{equation}
  \label{eq:schrodi}
  \partial_v^2\psi(v)= - \frac{\mathrm{e}^{2\phi_0}}{2}\,
  T_{vv}(v)\psi(v)\;. 
\end{equation}
Note that $\psi(v)$ is defined up to a multiplicative constant. 
Now, one can solve for $\psi(v)$ given the initial  data
$T_{vv}(v)$. After that the entire solution is determined by
Eq.~(\ref{eq:13}) and expressions from the previous Section. For
example, the outgoing energy flux equals 
\begin{equation} 
  \label{eq:16}
  T_{uu} (u) = \left(\lambda e^{\phi_0}\psi/ \partial_v\psi \right)^4\,
    T_{vv} \Big|_{v = V(u)}\;,
\end{equation}
where $V(u)$ is inverse of $U(v)$, $V(U(v)) = v$. We obtained
Eq.~(\ref{eq:16}) by substituting the reflection
law~(\ref{eq:riccati}) into the definition~(\ref{eq:7}) of the flux and then
expressing the derivative of $U(v)$ from the first of
Eqs.~(\ref{eq:riccati}) and Eq.~(\ref{eq:13}).

Importantly, Eq.~(\ref{eq:schrodi}) is well-known in
mathematical physics. Similar equation appears in
Liouville theory at classical and semiclassical
levels~\cite{Teschner:2010je}. Besides, the eigenstates of the Gaudin
model~\cite{Gaudin:1976sv} can be related to the solutions of
Eq.~(\ref{eq:schrodi}) with monodromies $\pm 1$ and rational
$T_{vv}(v)$~\cite{Feigin:1994in}. In what follows we exploit these
similarities for studying exact solutions in dilaton gravity.

The function $\psi(v)$ in Eq.~(\ref{eq:13}) has 
simple geometric meaning. First, the value of $\psi$ is related to the
proper time 
$\tau$ along the boundary, 
\begin{equation}
  \label{eq:17}
  d\tau^2=e^{2\phi_0}dU(v)\, dv=(\partial_v \psi/\lambda\psi)^2
  dv^2\qquad\Rightarrow\quad \psi(v) = \psi_0\cdot \mathrm{e}^{\lambda \tau(v)}\;,
\end{equation}
where we used Eqs.~(\ref{eq:6}),~(\ref{eq:riccati}),
(\ref{eq:13})  and introduced the arbitrary
constant $\psi_0$ related to the  origin of $\tau$. Function $\tau(v)$ is
illustrated in Fig.~\ref{fig:coordinates}.
Second, recall that $v$ is the exponent of the flat light-cone coordinate
$(t+u)$ far away from the boundary, Eq.~(\ref{eq:15}). Thus,
$\psi(v)$ maps the affine coordinate at $J^-$ to
$\tau$. Equation~(\ref{eq:schrodi}) relates this 
coordinate-independent function to the asymptotic Cauchy data
$T_{vv}(v)$.  

\begin{figure}[htb]

  \vspace{-13mm}
\centerline{\begin{minipage}[h]{0.36\linewidth}
\center{\includegraphics[width=0.99\linewidth]{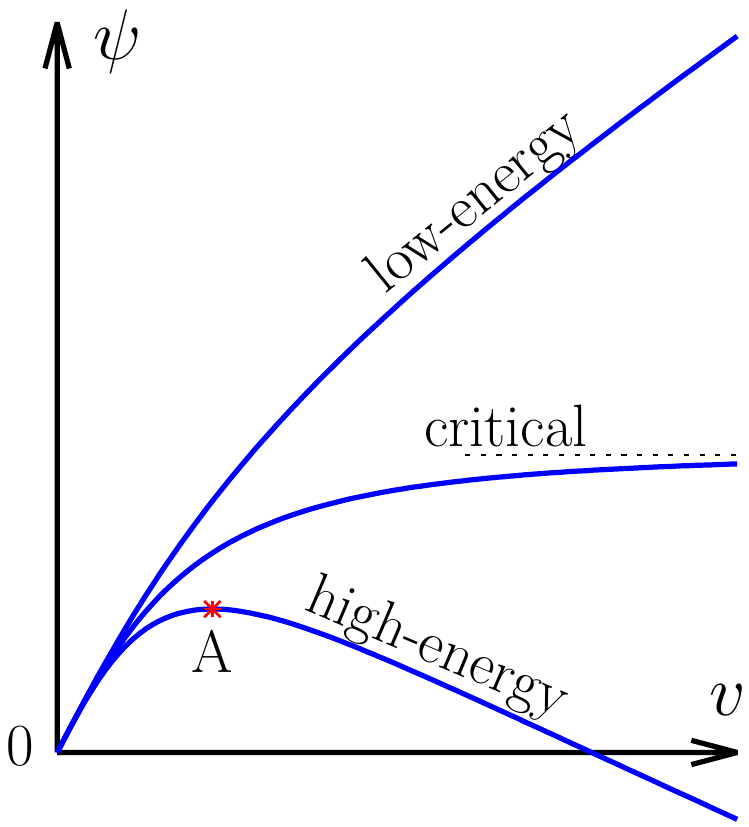}}
\end{minipage}
\hspace{10mm}
\begin{minipage}[h]{0.405\linewidth}
\center{\includegraphics[width=0.99\linewidth,angle=45]{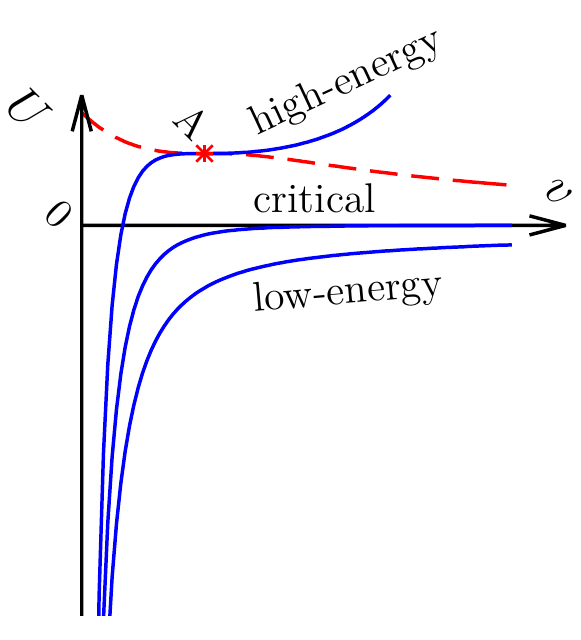}}
\end{minipage}}

\vspace{-8mm}
\hspace{3.2cm}(a) \hspace{8cm}(b)
\caption{Functions $\psi(v)$ and $U(v)$ at different $T_{vv}$. The
  right graph is rotated for visualization 
  purposes. Dashed line in this graph is the space-like ``singularity''
  $\phi = \phi_0$. \label{fig:psi}}
\end{figure}

Consider general properties of classical solutions in the model with
a boundary. Expression~(\ref{eq:17}) implies that $\psi(v)$ vanishes in
the infinite past,
\begin{equation}
  \label{eq:18}
  \psi(0) = 0\;.
\end{equation}
Indeed, behavior $\psi \to c_0 v$ as $v\to 0$ corresponds to the
linear dilaton vacuum (\ref{eq:14}) in the beginning of the
process. To simplify the next argument, we set\footnote{Recall that
  $\psi(v)$ is defined up to a multiplicative constant.} $c_0 =
1$. We consider well-localized $T_{vv}(v)$ and therefore linear 
asymptotics 
\begin{equation}
  \label{eq:19}
  \psi(v) \to C v + D \qquad \qquad\mbox{as} \qquad v\to +\infty
\end{equation}
of the solution to Eq.~(\ref{eq:schrodi}).
If $T_{vv}$ is small, one has $C
\approx 1$. The respective ``low-energy'' solutions describe reflection of 
matter waves off the time-like boundary, see Figs.~\ref{fig:psi}a,b.  As 
$T_{vv}$ grows,  the function $\psi(v)$ becomes more concave and
$C$ decreases because $\partial_v^2
\psi \propto -T_{vv} < 0$. For some large fine-tuned $T_{vv}(v)$ one
obtains critical solutions with $C = 0$. In this case the boundary is
null in the asymptotic future because its proper time $\tau(v)$ in
Eq.~(\ref{eq:17}) remains finite as $v\to +\infty$. 
The respective ``critical'' solution in Figs.~\ref{fig:psi} is at
the brink of black hole formation: we will see that the asymptotically
null boundary sits precisely at the horizon of would-be black hole.

At sufficiently high energies we get $C<0$ and therefore $\psi(v)$ has
a maximum (point $A$ in Fig.~\ref{fig:psi}a). The boundary is null at
this point: $dU/dv|_A  \propto  (\partial_v \psi)^2|_A {= 0}$
according to
Eqs.~\eqref{eq:riccati} and \eqref{eq:13}. Moreover,
near $A = (u_A,\, v_A)$ one obtains ${U(v) \approx u_A + d\cdot
(v-v_A)^3}$ and 
\[
\mathrm{e}^{-2\phi(u,\,v)} \approx \mathrm{e}^{-2\phi_0} +
  \frac{T_{vv}(v_A)}{4 d^{2/3}} \left[ (u_A-u)^{2/3} -  d^{2/3}(v - v_A)^2 \right] \;.
\]
where Eqs.~(\ref{eq:riccati}), (\ref{eq:13}), and (\ref{eq:bulk_gravity})
were solved to the leading order in $u - u_A$, $v - v_A$. Thus, 
$A$ is a singularity of $\phi$ in coordinates $(u,\, v)$.

Besides, one discovers 
that the condition $\phi = \phi_0$ defines {\it two} intersecting
curves $u - u_A \approx \pm d (v - v_A)^3$ near $A$, and only one of
those is the time-like boundary considered so far. The second curve is
space-like, it is shown by the dashed line in Fig.~\ref{fig:psi}b. The
boundary conditions (\ref{eq:riccati}) are not met at this line.  We
obtained the analog of the black hole singularity in the model with a
boundary. Indeed, our 
model is formulated at $\phi < \phi_0$ i.e.\ in the space-time region
to the right of both solid and dashed graphs in
Fig.~\ref{fig:psi}b. The space-like ``edge'' $\phi = \phi_0$ swallows  
all matter at $u>0$ limiting the region accessible to the outside
observer to $u<0$. The line $u=0$ is a horizon.

Except for the point $A$ itself, the solution is smooth at
the space-like ``singularity'' $\phi = \phi_0$. This fact was not
appreciated in the previous studies. The
mass of the formed black hole, by energy conservation, is related  to
the value of the dilaton field at the future horizon,
$$
M_{bh} = \int\limits_0^{+\infty} \lambda v dv \, T_{vv}  -
  \int\limits_{-\infty}^{0} \lambda |u|du \, T_{uu} = 2\lambda\left[g(+\infty) +
  h(0)\right] = 
  2\lambda \lim\limits_{v\to +\infty}\,
  \mathrm{e}^{-2\phi(0,\, v)}\;,
$$
where we subtracted the final matter energy from the initial one
in the first equality (cf.\ Eq.~\eqref{eq:15}), integrated by parts and used
Eqs.~(\ref{eq:g_h_def}) in the second equality, and then expressed the
result in terms of 
$\phi$, Eq.~(\ref{eq:bulk_gravity}). Since $\phi < \phi_0$, this
implies that all black hole masses  are larger than
\begin{equation}
  \label{eq:23}
  M_{cr} = 2\lambda \mathrm{e}^{-2\phi_0}\;,
\end{equation}
see detailed discussion in~\cite{Russo:1992ax, Bose:1996pi}. Black
holes with $M_{bh} = M_{cr}$ have boundary sitting precisely at the
horizon. They  are formed in the critical solutions. 

The solutions in Fig.~\ref{fig:psi}b, when replotted in the 
finite-range coordinates $(\bar{u},\, \bar{v}) = (\mathrm{arctan}\,
u,\, \mathrm{arctan}\, v)$, look like Penrose diagrams, see
Fig.~\ref{fig:intro_solutions}. From 
now on, we will exploit $\bar{u}$  and $\bar{v}$ for visualizing the
solutions. We will also mark the (smooth) space-like ``singularities'' 
$\phi = \phi_0$ by zigzag lines, see the one in
Fig.~\ref{fig:intro_solutions}c.    

\subsection{On-shell conformal symmetry}
\label{sec:shell-conf-symm}
We find that the boundary equation~(\ref{eq:schrodi}) is invariant
under conformal transformations $v\to w(v)$,
\begin{align}\label{eq:conf_trans}
  & \psi\,\to\;\tilde{\psi}(w)=\left(\frac{dv}{dw}\right)^{-1/2}\psi(v)\;,\\
  & T_{vv}\,\mapsto\;\tilde{T}_{vv}(w)=\left(\frac{dv}{dw}\right)^{2}
    T_{vv}(v)+\mathrm{e}^{-2\phi_0}\{v;\,w\}\;,
\label{eq:21}
\end{align}
which change $\psi(v)$  as an $h=-1/2$ primary field and 
$T_{vv}(v)$ as an energy-momentum tensor with large negative central
charge $c = -24 \pi \mathrm{e}^{-2\phi_0}$~~\cite{DiFrancesco:1997nk}. In
Eq.~(\ref{eq:conf_trans}) we introduced the Schwarzian derivative $\{v;\,
w\} \equiv v'''/v' - 3(v'')^2/2(v')^2$ with ${v' \equiv dv/dw}$.
The transformations~(\ref{eq:conf_trans}), (\ref{eq:21})
relate physically distinct solutions\footnote{Unlike the
  transformations in~\cite{Das:1994kr, Strominger:1994xi}, they do not
  represent residual gauge symmetry. The latter was completely fixed,
  see discussion after Eq.~(\ref{eq:6}).} with 
different energy fluxes $T_{vv}$. Acting with them on the vacuum $\psi
= v$, $T_{vv} = 0$ one can obtain any solution. 

Note that the symmetry (\ref{eq:conf_trans}), (\ref{eq:21}) does not
make our model a CFT in a conventional sense\footnote{Thus, one may
  still hope that our model is unitary at the quantum level
  despite negative primary dimension in Eq.~(\ref{eq:conf_trans}) and
  negative central charge in Eq.~(\ref{eq:21}).}. First, the full
energy-momentum tensor $T_{\mu\nu}+T_{\mu\nu}^{(\phi)}$ of the model
includes contribution of the dilaton field and vanishes by Einstein
equations, cf.~Eq.~(\ref{eq:1}). Second,  Eq.~(\ref{eq:21}) is not a
conformal transformation $f_{in}\to f_{in}(v(w))$ of the massless
scalar field $f$ far away from  the boundary: the latter changes  
classical $T_{vv}\equiv (\partial_v f_{in})^2$    without the
Schwarzian derivative. At the quantum level, healthy 
conformal matter has positive central charge
$c>0$~\cite{DiFrancesco:1997nk}, and transformations of its 
energy-momentum tensor $T_{vv}$ do not match Eq.~(\ref{eq:21}) as
well.

Transformations from the $\mbox{SL}(2,\, \mathbb{R})$ subgroup
of (\ref{eq:conf_trans}), (\ref{eq:21}),
\begin{equation} 
  \label{eq:22}
v\to  w(v) = \frac{\alpha v+\beta}{\gamma v+\delta}\;, \qquad\qquad
\alpha \delta - \beta \gamma=1\;,
\end{equation}
have vanishing Schwarzian derivative and therefore change $f$ in the
standard way ${f_{in}\to   f_{in}(v(w))}$. Besides trivial
translations of $v$ they include $v$-dilatations due to shifts of the
asymptotic coordinate $t+x$ in Eq.~(\ref{eq:15}) and inversion $v\to
1/v$ related to PT-reflection $t+x \to -(t+x)$.
These transformations 
constitute the global symmetry group of our model.

As a side remark, let us argue that~(\ref{eq:conf_trans}),
(\ref{eq:21}) is a symmetry of the gravitational degrees of
freedom but not of the matter sector. To this end we introduce the field $\chi(u) =
\mathrm{e}^{-\lambda \tau(u)}/\psi_0$ which is $T$-symmetric with
respect to $\psi(v)$   and therefore satisfies
\begin{equation} 
  \label{eq:24}
  \partial_u^2 \chi(u) = - \frac{\mathrm{e}^{2\phi_0}}{2} \, T_{uu}(u)
  \chi(u)\;,
\end{equation}
cf.~Eqs.~(\ref{eq:17}) and~(\ref{eq:schrodi}); now, $\tau(u)$ is
the boundary proper time parametrized with~$u$. It is convenient to 
combine $\psi(v)$ and $\chi(u)$ into a single free field  
$$
\mathrm{e}^{-2\Phi(u,v)} \equiv \chi(u)\psi(v)
\,\mathrm{e}^{-2\phi_0}\;,
$$
transforming in a simple Liouville-like manner
under Eq.~(\ref{eq:conf_trans}). To describe the gravitational degrees
of freedom with $\Phi$, we extract its
energy-momentum tensor $T_{\mu\nu}^{(\phi)}$ from the Einstein
equations ${T_{\mu\nu}^{(\phi)} +  T_{\mu\nu} = 0}$,
$$  T_{vv}^{(\phi)} \equiv - T_{vv} = 8\mathrm{e}^{-2\phi_0} \left[(\partial_v
    \Phi)^2 - \partial_v^2 \Phi/2\right]\;, \qquad T_{uu}^{(\phi)}
  \equiv - T_{uu}\;,
$$
where Eq.~(\ref{eq:schrodi}) was used in the left equality;  similar
expression for $T_{uu}^{(\phi)}$ can be obtained using Eq.~(\ref{eq:24}). One
observes that $T_{vv}^{(\phi)}$ transforms under
Eq.~(\ref{eq:conf_trans}) as an energy-momentum tensor with positive 
conformal charge $c = 24\pi \mathrm{e}^{2\phi_0}$, in agreement
with Eq.~(\ref{eq:21}).

Now, the entire scattering problem can be reformulated in terms of
$\Phi$. One sends the incoming energy flux  
$T_{vv}^{(\phi)}$ towards the dynamical boundary $u = U(v)$ at  $\Phi
= \phi_0$. The flux reflects into $T_{uu}^{(\phi)}$ according to the
non-conformal law $T_{uu}^{(\phi)} = (dU/dv)^{-2}\, T_{vv}^{(\phi)}$, 
see Eq.~(\ref{eq:16}). All these equations and boundary conditions  can be
summarized in the flat-space action 
$${S_\Phi= -\int_{\Phi<\phi_0}
  d^2x\,[\mathrm{e}^{-2\phi_0}(\partial_\mu\Phi)^2+\lambda^2]}\;.
$$
In this setup (\ref{eq:conf_trans}), (\ref{eq:21}) is an apparent
conformal symmetry of $\Phi$ far away from the boundary, 
whereas the symmetry of the matter sector is hidden in the reflection laws.
\begin{sloppy}

\section{Integrable sector}\label{sec:int_sector}
\subsection{General solution}
\label{sec:general-solution}
One can use  Eq.~(\ref{eq:schrodi}) to express the entire solution  in
terms of one arbitrary function. Indeed, introducing
\begin{equation} 
  \label{eq:35}
  W\equiv \partial_v \psi/\psi = \mathrm{e}^{2\phi_0} \left(
  \partial_v g - \lambda^2 U\right)\;,
\end{equation}
we find,
\begin{equation}
  \label{eq:34}
  \psi=e^{\int\limits^v dv'\,W(v')}\;, \qquad
  -\frac{\mathrm{e}^{2\phi_0}}{2}T_{vv}= W^2+\partial_v W\;. 
\end{equation}
Then $U$,  $T_{uu}$, $\phi$, and $f$ are given by Eqs.~(\ref{eq:35}),
(\ref{eq:16}),  (\ref{eq:bulk_gravity}),  and~(\ref{eq:7}). We obtained
general classical solution in the model with a boundary.

By itself, this solution is of little practical use because the
function $\psi(v)$ has a zero at $v=0$ and, possibly, another one at $v
= \tilde{v}_1 > 0$, see Fig.~\ref{fig:psi}a. In general, the incoming
flux $T_{vv}(v)$ in Eq.~(\ref{eq:34}) is singular at these
points. Indeed, Eq.~(\ref{eq:35}) gives
$$
W(v) = R(v) + 1/v + 1/(v - \tilde{v}_1)\;,
$$
where $R(v)$ is regular at $v\geq 0$. As a consequence, 
$T_{vv}(v)$ has first-order poles  at $v = 0$ and
$\tilde{v}_1$. Requiring zero residuals at these poles, we obtain two
constraints  $R(0) = -R(\tilde{v}_1) = 1/\tilde{v}_1$ on parameters of
$R(v)$. 

Choosing multiparametric $R(v)$ and solving the constraints, one finds
an arbitrary number of smooth solutions. The physical
ones satisfy
\begin{equation}
  \label{eq:37}
  T_{vv}(v)\geq 0 \;, \qquad \mbox{at}\qquad v\geq 0\;.
\end{equation}
In what follows we will concentrate on a large class of
soliton solutions with power-law singularities. We will
argue that some of them satisfy Eq.~(\ref{eq:37}). 

\end{sloppy}
\subsection{Soliton solutions with power-law singularities}
\label{sec:sing-solut}
Let us follow the Painlev\'e test~\cite{Vernov:2002rw} and guess the
form of $T_{vv}(v)$ which guarantees that the general solution
$\psi(v)$ of Eq.~(\ref{eq:schrodi}) has power-law singularities ${\psi
  \sim (v - v_0)^{-s}}$ in the complex $v$-plane. One introduces 
Laurent series at $v \approx v_0$,
\begin{equation}
  \label{eq:28}
  -\frac{\mathrm{e}^{2\phi_0}}{2}\, T_{vv} = \sum_{k=0}^{+\infty}
    T_{k-2} (v - v_0)^{k-2} \;, \qquad \qquad   \psi =
    \sum_{k=0}^{+\infty} \psi_{k-s} (v-v_0)^{k-s}\;, 
\end{equation}
where the expansion of $T_{vv}$ starts from $(v-v_0)^{-2}$ due to
Eq.~(\ref{eq:schrodi}). Substituting Eqs.~(\ref{eq:28})
into Eq.~(\ref{eq:schrodi}), we obtain an infinite algebraic system
for $\psi_{k-s}$, 
\begin{equation}
  \label{eq:29}
  (k-s)(k-s-1) \psi_{k-s} = T_{-2} \psi_{k-s} + T_{-1} \psi_{k-s-1} +
  \dots + T_{k-2} \psi_{-s}\;.
\end{equation}
The very first ($k=0$) of these relations gives $T_{-2} = s(s+1)$, the others
determine $\psi_{k-s}$ with $k\geq 1$ in terms of arbitrary $\psi_{-s}$
and $\{T_m\}$. Expression~(\ref{eq:28}) is a  general solution
of the second-order equation~(\ref{eq:schrodi}) if precisely two
of its parameters, $\psi_{-s}$ and some $\psi_{k_0-s}$, remain
arbitrary. Thus, $(k_0-s)(k_0-s-1) = s(s+1)$  in Eq.~(\ref{eq:29})
implying $k_0 = 2s+1$. One concludes that  $s$ is integer or half-integer.

Note that the two equations from the system~(\ref{eq:29}) which do not
determine the coefficients of $\psi$, constrain  $\{ T_{k}\}$. For
example for $s=1/2$ one gets,
\begin{equation}
  \label{eq:31}
  T_{-2} = 3/4\;, \qquad \qquad T_0 = (T_{-1})^2 \;,
\end{equation}
where we expressed all $\psi_{k-1/2}$ via  $\{T_k\}$ and
$\psi_{-1/2}$. For larger $s$, one obtains $T_{-2} = s(s+1)$ and
higher-order equations listed in Table~\ref{tab:equations}.
\begin{table}[htb]
\centerline{\begin{tabular}{l|l}
$s$ & equation\\
\hline
$1$ & $T_1 = T_0 T_{-1} - \frac14 (T_{-1})^3$\\
$3/2$ & $ T_2  = \frac23 T_1 T_{-1} - \frac{5}{18} T_0
(T_{-1})^2 + \frac14 (T_0)^2 + \frac1{36} (T_{-1})^4$\\
$2$ & \dots
\end{tabular}}
\caption{Equations for the Laurent coefficients of the solitonic
  $T_{vv}(v)$.\label{tab:equations}}
\end{table}

We arrived at the practical method for obtaining the soliton solutions
in our model. One specifies $N$ singularities 
of $\psi(v)$: selects their integer or half-integer powers $s_n$ and
complex positions $v_n$. The function $T_{vv}(v)$ has second-order
poles at $v = v_n$, see
Eq.~(\ref{eq:28}). This analytic structure gives expressions,
\begin{equation}
  \label{eq:32}
  -\frac{\mathrm{e}^{2\phi_0}}{2} \, T_{vv} = \sum\limits_{n=1}^{N}
  \left[ \frac{s_n(s_n+1)}{(v - v_n)^2} + \frac{T_{-1}^{n}}{v -
      v_n}\right]\;,\qquad 
    \psi = C\,  \frac{\prod_{m=1}^M (v - \tilde{v}_m)}{\prod_{n=1}^{N}(v -
    v_n)^{s_n}} \;,
\end{equation}
where we required $T_{vv} \to 0$ as $v\to +\infty$ and introduced a
polynomial in the nominator of $\psi(v)$ with $M$ zeroes $\tilde{v}_m$
and a
normalization constant  $C$. Next, one solves
equations in Table~\ref{tab:equations} at each singularity and
determines $T_{-1}^n$. After that $\psi(v)$ is obtained by
substituting Eqs.~(\ref{eq:32}) into 
Eqs.~(\ref{eq:schrodi}) or~(\ref{eq:29}). Two parameters~---  say, $C$
and   $\tilde{v}_M$~--- remain arbitrary because Eq.~(\ref{eq:32}) is
a general solution of the second-order equation. One  takes 
$\tilde{v}_M=0$ in accordance with the flat-space
asymptotics~(\ref{eq:18}). This gives the soliton $\{
\psi(v),\, T_{vv}(v)\}$ characterized by $N$ complex parameters $v_n$
and the same number of integers or half-integers~$s_n$. 
\begin{sloppy}

We consider solutions with finite total energy of incoming matter, 
$$
E_{in} = \int_0^{+\infty} \lambda v dv \, T_{vv}(v)\;,
$$
see Eq.~(\ref{eq:15}). Convergence of this integral
implies ${T_{vv} \sim \bar{o}(v^{-2})}$ as ${v\to +\infty}$ or, given
Eq.~(\ref{eq:32}), linear relations
\begin{equation}\label{eq:finiteness} 
  \sum_{n=1}^N T_{-1}^n = 0\;, \qquad\qquad
  \sum_{n=1}^N \left[s_n(s_n+1)+v_n T_{-1}^n \right]=0\;. 
\end{equation}
Moreover, asymptotic (\ref{eq:19}) of $\psi(v)$ suggests
falloff ${T_{vv} \sim O(v^{-4})}$ at large $v$ and  additional relation
\begin{equation}\label{eq:add_constraint}
\sum_{n=1}^N\left[2v_n s_n(s_n+1)+v_n^2T_{-1}^n\right]=0\;,
\end{equation}
which should hold for noncritical  solutions. Equations
(\ref{eq:finiteness}) and (\ref{eq:add_constraint}) are useful for
obtaining the lowest solitons. 

{\bf Example.} Consider the soliton with two $s=1/2$
singularities\footnote{Note that $T_{vv}(v)$ with one
  singularity  does not satisfy Eqs.~(\ref{eq:finiteness}).}. Solving
the finite-energy    
conditions~(\ref{eq:finiteness}), one obtains $T_{-1}^1 = - T_{-1}^2 =
3/[2(v_2 -v_1)]$. It is straightforward to check that $T_{vv}(v)$
with these parameters satisfies Eqs.~(\ref{eq:31}) at $v = v_1$ and $v=
v_2$. To make the solution real at $v\in \mathbb{R}$, we take $v_1 =
a+ib$ and $v_2 =a-ib$. Then Eqs.~(\ref{eq:32}) give,
\begin{equation}
  \label{eq:33}
  T_{vv} = \frac{6\mathrm{e}^{-2\phi_0}\, b^2}{[(v-a)^2 + b^2]^2}\;,
  \qquad\qquad
  \psi(v) = \frac{v( a^2+b^2 - av)}{[(v-a)^2 + b^2]^{1/2}}\;,
\end{equation}
where $\psi(v)$ was obtained by substituting Eqs.~(\ref{eq:32})
into Eq.~(\ref{eq:schrodi}). One observes that the matter flux
(\ref{eq:33}) peaks near $v\sim a$, its total energy ${E_{in}=
  \frac{3}{2}   M_{cr}   \left[1+  \frac{a}{b} \,
    \mathrm{arcctg}(-a/b)\right]}$ is controlled by the ratio $a/b$  ,
where   $M_{cr} = 2\lambda \mathrm{e}^{-2\phi_0}$ is the minimal black
hole mass.

\end{sloppy}
\begin{figure}[htb]

  \vspace{-9mm}
  \centerline{
    \hspace{-0.07\textwidth}
    \includegraphics[width=0.28\textwidth,angle=45]{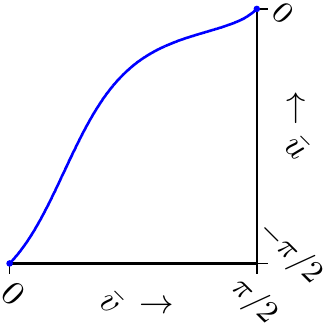}
    \hspace{-0.09\textwidth}
    \includegraphics[width=0.28\textwidth,angle=45]{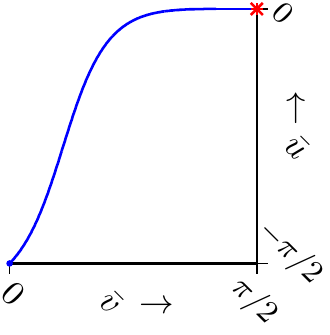}
    \hspace{-0.09\textwidth}
    \includegraphics[width=0.28\textwidth,angle=45]{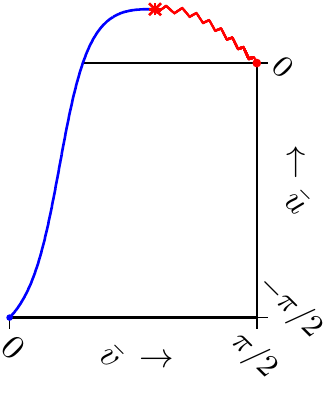}}

  \vspace{-3mm}
  \hspace{5mm}(a) $\lambda a = -0.2$ \hspace{23mm} (b)
  $\lambda a=0$
  \hspace{3.1cm}(c) $\lambda a=0.2$ 
  \caption{Solution (\ref{eq:33}) in the finite-range coordinates
    $\bar{u} = \mathrm{arctg}(\lambda u)$, $\bar{v} =
    \mathrm{arctg}(\lambda v)$ at different values of $a$. We use
    $\lambda b = \mathrm{e}^{2\phi_0} = 1$ keeping in mind that the
    parameter $\mathrm{e}^{2\phi_0} \ll 1$ can be 
    restored in the classical solution, see discussion in 
    Sec.~\ref{sec:adding-boundary}. \label{fig:halfpole}}
\end{figure}
Since $\psi \to -av$ as $v\to +\infty$, the solution~(\ref{eq:33})
describes reflection of matter waves off the boundary and formation of
black holes at $a<0$ and $a>0$, respectively, see
Fig.~\ref{fig:psi}a. This fact is clearly seen in 
Fig.~\ref{fig:halfpole} showing the boundary $u = U(v)$ at different
$a$ in the finite-range coordinates 
$(\bar{u},\, \bar{v})$. In Fig.~\ref{fig:halfpole}c we also
plotted the space-like ``singularity'' $\phi = \phi_0$ and horizon
$u=0$ (zigzag red and solid black lines, respectively). Note that the
critical solution in Fig.~\ref{fig:halfpole}b corresponds to $E_{in} =
\frac32 M_{cr}$.

\begin{figure}[htb]
  \centerline{\includegraphics[width=0.37\textwidth]{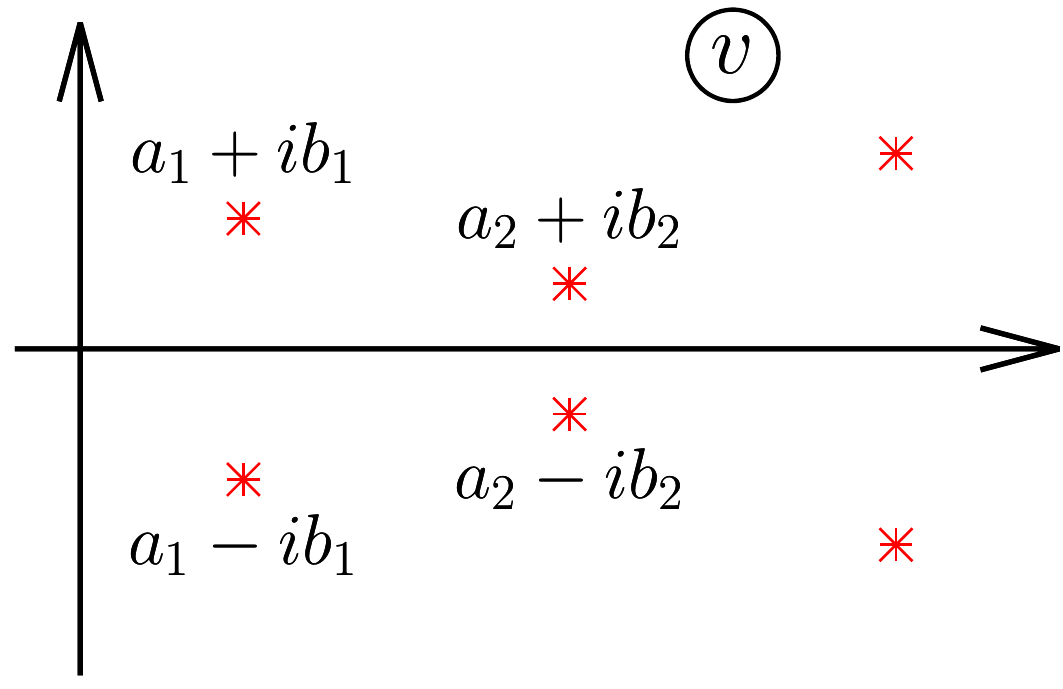}}
  \caption{Singularities of solitons in the complex
    $v$-plane.\label{fig:sing}}
\end{figure}
The simplest exact solution in Eq.~(\ref{eq:33}) describes the incoming
matter flux with a single peak. Solutions with multiple peaks  can be
obtained by adding  singularities at $v=a_n\pm ib_n$, see
Fig.~\ref{fig:sing}. Unfortunately, it is hard to find these solutions
explicitly at large $N$. Besides, it is not clear whether they will satisfy
the positivity condition~(\ref{eq:37}). We will clarify these issues
in the subsequent Sections.

\subsection{Simplifying the coefficient equations}
\label{sec:simpl-algebr-equat}
Instead of solving the equations in Table~\ref{tab:equations}, one can
extract $T_{vv}(v)$ from the general solution. Namely, substituting the
solitonic $\psi(v)$ into the first of Eqs.~(\ref{eq:34}), we find,
\begin{equation}
  \label{eq:new_notation}
W(v)=-\sum_{n=1}^N\frac{s_n}{v-v_n}+\sum_{m=1}^M\frac1{v-\tilde{v}_m}\;.
\end{equation}
Then the second of Eqs.~(\ref{eq:34}) gives the incoming
flux. However, in this case  
$T_{vv}(v)$ receives parasitic first-order poles at 
$v=\tilde{v}_m$ which are absent in Eq.~(\ref{eq:32}). Requiring zero
residuals at these poles, we obtain equations for $\{\tilde{v}_m \}$, 
\begin{equation}
  \label{eq:condition_w}
  \sum_{n=1}^N\frac{s_n}{\tilde{v}_m-v_n}=\sum_{\substack{m'=1 \\ m'\ne
    m}}^M\frac1{\tilde{v}_m-\tilde{v}_{m'}}\;,
\end{equation}
which are, in fact, equivalent to the ones in
Table~\ref{tab:equations}. Indeed, after solving Eqs.~(\ref{eq:condition_w})
one obtains $T_{vv}(v)$ of the form (\ref{eq:32}) with
\begin{equation} 
  \label{eq:30}
  T_{-1}^n=\sum_{n'\ne n}\frac{2s_n
          s_{n'}}{v_n-v_{n'}}-\sum_{m}\frac{2s_n}{v_n-\tilde{v}_m}\;.
\end{equation}
In practice one finds $\tilde{v}_m$ numerically from
Eqs.~(\ref{eq:condition_w}), then computes $T_{vv}$ and $\psi$ 
by Eqs.~(\ref{eq:30}) and (\ref{eq:32}).

Unlike in Sec.~\ref{sec:general-solution}, we impose
Eqs.~(\ref{eq:condition_w}) at all $\tilde{v}_m$, not just the ones at
the real positive axis.  The goal is to obtain solutions with transparent
properties, see the forthcoming discussion in Sec.~\ref{sec:relat-gaud-model}. 

\subsection{$\mathrm{SL}(2,\, \mathbb{C})$ symmetry}
\label{sec:mboxsl2--mathbbr}
The global $\mbox{SL}(2,\, \mathbb{C})$ transformations~(\ref{eq:22})
are invertible and therefore preserve the singularity structure of the
solitons. One obtains,
\begin{equation}
  \label{eq:38}
 T_{vv} \to \tilde{T}_{vv}(w) = \frac{T_{vv}(v)}{(\alpha - \gamma w)^4}\;,
   \qquad 
   \psi \to \tilde{\psi}(w) = (\alpha -  \gamma w) \, \psi(v) \;.
\end{equation}
This symmetry relates solitons with different parameters. Real
solutions at $v\geq 0$ transform under $\mbox{SL}(2,\, \mathbb{R})$.
 
The transformation (\ref{eq:22}) sends the point $v = -\delta /\gamma$ to
infinity. If the original solution was regular at this point, its image
receives asymptotics $\tilde{\psi} \to  Cw + D$ and
${\tilde{T}_{vv} \to O(w^{-4})}$ as $w\to +\infty$. In
Eq.~(\ref{eq:19}) we obtained the same
asymptotics from physical considerations. Solutions with  other
asymptotics, i.e.\ those violating the finite-energy
conditions~(\ref{eq:finiteness}) or Eq.~(\ref{eq:add_constraint}),
have singularities ``sitting'' at infinity.

{\bf Example.} One can use the above property to construct new
solutions. Consider e.g.\ the trivial solution $\psi =
v^{-s} - v^{s+1}$,  $T_{vv} = -2\mathrm{e}^{-2\phi_0} \, 
s(s+1)/v^{2}$ of Eq.~(\ref{eq:schrodi}) with non-linear $\psi(v)$ at
large $v$. We send the points $v=0,\, \infty$, and~$1$ to $v_1$,
$v_2$, and $0$ by linear rational transformation\footnote{With
  parameters   $\alpha = -\beta = (1/v_2   - 1/v_1)^{-1/2}$, $\gamma =
  \alpha/v_2$,   $\delta  = \beta/v_1$.} 
(\ref{eq:38}) and get,
\begin{equation}
  \label{eq:40}
  T_{vv} = \frac{-2\mathrm{e}^{-2\phi_0} s(s+1) (v_2 -
    v_1)^2}{(v  - v_1)^2 (v - v_2)^2}\;, \qquad
    \psi = \frac{i(v - v_1)^{s+1}v_2^s}{v_1^{s+1}(v - v_2)^{s}}
  - \frac{i(v - v_2)^{s+1}v_1^s}{v_2^{s+1}(v - v_1)^{s}}\;, 
\end{equation}
where the constant in front  of $\psi(v)$ was ignored. This is the 
soliton with two singularities of power $s$. Taking $v_1 = v_2^*
= a+ib$, one obtains $T_{vv}(v)\geq 0$ at real  $v$. Note that 
the incoming flux in Eq.~(\ref{eq:40}) is the same as in
Eq.~(\ref{eq:33}) albeit with different multiplicative factor. The
behaviors of the boundaries are also similar, as one 
can see by comparing the  solutions~(\ref{eq:40}) with\footnote{In  
  Figs.~\ref{fig:intro_solutions}a, b, and c we used $\lambda a = -1$,
  $-1/\sqrt{3}$, and $0.3$, respectively, and $\lambda b   =
  \mathrm{e}^{-2\phi_0} =  1$.}
$s = 1/2$ and $1$ in Figs.~\ref{fig:halfpole}
and~\ref{fig:intro_solutions}, respectively.

\subsection{Relation to the Gaudin model}
\label{sec:relat-gaud-model}
In this Section we establish one-to-one correspondence between the
solitons (\ref{eq:32}) and eigenstates of the
auxiliary integrable system, the Gaudin model~\cite{Gaudin:1976sv,
  Feigin:1994in, Frenkel:1995zp}. This will allow us to count the
number of solitons and explain some of their properties.

\begin{sloppy}
The Gaudin model~\cite{Gaudin:1976sv}
describes a chain of $N$ three-dimensional spins
${\hat{\boldsymbol{s}}_n = \{\hat{s}_n^1,\, \hat{s}_n^2,\, \hat{s}_n^3\}}$ with the
standard commutation relations $[\hat{s}_n^\alpha,\,   \hat{s}_l^\beta
] = i\delta_{nl}\, \epsilon^{\alpha\beta\gamma} \hat{s}_n^\gamma$.
The model is equipped with $N$  commuting Hamiltonians 
\begin{equation}
  \label{eq:gaudin_hamiltonian}
  \hat{\cal T}_n=\sum_{l\ne
    n}\frac{(\hat{\boldsymbol{s}}_n,\, \hat{\boldsymbol{s}}_l)}{v_n-v_l}\;,  
\end{equation}
where  $v_n$ are complex parameters and $(\hat{\boldsymbol{s}}_n,\,
\hat{\boldsymbol{s}}_l) \equiv \sum_{\alpha}\hat{s}_n^\alpha\,  \hat{s}_l^{\alpha}$ is
the scalar product. The eigenstates $|\Psi\rangle$ of the model
simultaneously diagonalize all Hamiltonians,
  ${\hat{\cal T}_n |\Psi \rangle = {\cal T}_n |\Psi\rangle}$, where
  ${\cal T}_n$ are complex eigenvalues. 

\end{sloppy}
It is convenient to pack all spins and Hamiltonians into
the operator-valued functions 
\begin{equation}
  \label{eq:T_operator}
\hat{\boldsymbol{s}}(v)\equiv \sum_{n=1}^N\frac{\hat{\boldsymbol{s}}_n}{v-v_n}
\;,\qquad
\hat{\cal T}(v) \equiv \left[\hat{\boldsymbol{s}}(v)\right]^2 = 
\sum_{n=1}^N\left[\frac{\hat{\boldsymbol{s}}_n^2}{(v-v_n)^2}+\frac{2\hat{\cal
      T}_n}{v-v_n}\right]\;.
\end{equation}
Now, the eigenvectors satisfy $\hat{\cal T}(v)
|\Psi\rangle= {\cal T}(v)|\Psi\rangle$. 

A complete set of eigenvectors and eigenvalues in the model
(\ref{eq:gaudin_hamiltonian}) is provided by the algebraic Bethe
Ansatz~\cite{Gaudin:1976sv,   Feigin:1994in, Frenkel:1995zp}. We
review this method in Appendix~\ref{sec:gaudin-spin-chain} and list
its main results below.

One fixes the representations 
$(\hat{\boldsymbol{s}}_n)^2 = s_n (s_n+1)$ of all spins, where $s_n$
are integers or  half-integers. The simplest eigenstate $|0\rangle$ of the Gaudin
model has all spins down,  
\begin{equation}\label{eq:gaudin_vacuum}
\hat{s}^-_n|0\rangle=0\;, \qquad
\hat{s}_n^3|0\rangle=-s_n|0\rangle \qquad \mbox{for all $n$}\;,
\end{equation}
where $\hat{s}_n^{-}\equiv \hat{s}_n^1-i\hat{s}_n^2$ are the lowering operators. The
other eigenstates are obtained by acting on $|0\rangle$ with
rising operators $\hat{s}^{+}(v) \equiv \hat{s}^{1}(v) + i\hat{s}^{2}(v)$,
\begin{equation}
  \label{eq:bethe_states}
|\tilde{v}_1,\, \dots ,\, \tilde{v}_M \rangle =
\hat{s}^+(\tilde{v}_1)\hat{s}^+(\tilde{v}_2)\dots 
  \hat{s}^+(\tilde{v}_M)|0\rangle
\end{equation}
at certain points $\tilde{v}_m$ which satisfy the Bethe equations,
\begin{equation}
  \label{eq:41}
-\sum_{n=1}^N\frac{s_n}{\tilde{v}_m-v_n} + \sum_{\substack{m'=1 \\ m'\ne
    m}}^M\frac1{\tilde{v}_m-\tilde{v}_{m'}} = 0\;.
\end{equation}
The eigenvalue of $\hat{\cal T}(v)$ corresponding to the state 
(\ref{eq:bethe_states}) is
\begin{equation}
  \label{eq:39}
        {\cal T}(v) = W^2 + \partial_v W\;, 
        \qquad \qquad W(v) =-  \sum_{n=1}^{N} \frac{s_n}{v - v_n} +
        \sum_{m=1}^{M} \frac{1}{v - \tilde{v}_m}\;.
\end{equation}
To sum up, one solves Eqs.~(\ref{eq:41}) for every $M$ and finds all
$\prod_n (2s_n+1)$ eigenvectors and eigenvalues of $\hat{\cal T}(v)$.    

\begin{sloppy}
Importantly, the Bethe equations~(\ref{eq:41}) coincide with the
algebraic equations~(\ref{eq:condition_w}) for the parameters
$\tilde{v}_m$ of the solitons in dilaton
gravity. This establishes one-to-one correspondence between 
our exact solutions and the eigenstates ~(\ref{eq:bethe_states}) of the 
Gaudin model. The singularities $\{s_n,\, v_n\}$ and zeros
$\{\tilde{v}_m\}$ of $\psi(v)$ are related to the parameters of the  
Gaudin Hamiltonians~(\ref{eq:gaudin_hamiltonian}) and Bethe
states~(\ref{eq:bethe_states}), respectively. Besides, the incoming flux
$T_{vv}(v)$ is proportional to the eigenvalue of $\hat{\cal   T}(v)$:
$T_{vv}(v) = -2\mathrm{e}^{-2\phi_0} \, {\cal T}(v)$,
cf.\ Eqs.~(\ref{eq:34}), (\ref{eq:new_notation}) and (\ref{eq:39}). 
The related quantities of the two models are listed in
Table~\ref{tab:relation}.

\end{sloppy}
\begin{table}[htb]
  \centerline{\begin{tabular}{c| l|l}
    & Solitons & Eigenstates of the Gaudin model\\
    \hline
    $v_n$ & positions of singularities& parameters of the
    Hamiltonians \\
    $s_n$ & powers of singularities & representations of
    $\hat{\boldsymbol{s}}_n$\\ 
    $\tilde{v}_m$ & zeros of $\psi(v)$ & parameters of  
    eigenstates \\
    $T_{vv} = -2\mathrm{e}^{-2\phi_0}  {\cal T}$ & incoming
    energy flux & eigenvalue of 
    $\hat{\cal T}(v)$\\
    $T_{-1}^n = 2{\cal T}_n$ & coefficients of $T_{vv}$ & eigenvalues
    of $\hat{\cal T}_n$\\ 
    \hline
  \end{tabular}}
  \caption{Correspondence between solitons in
    dilaton gravity and eigenstates of the Gaudin
    model.\label{tab:relation}}
  \end{table}

One can use the Gaudin model to study solitons in dilaton
gravity. We are interested in the case of well-localized
$T_{vv} = O(v^{-4})$ as ${v\to +\infty}$. The
corresponding Gaudin states have zero total spin\footnote{Note that 
  $\hat{\boldsymbol{S}}$ commutes with all Gaudin Hamiltonians.}
$$
\hat{\boldsymbol{S}} = \sum_n \hat{\boldsymbol{s}}_n
$$
because $\hat{\cal T} \to (\hat{\boldsymbol{S}}/v)^2$ as $v\to
+\infty$, see Eq.~(\ref{eq:T_operator}). Using this property, one 
counts the number of solitons with correct
asymptotics by adding up spins. For example, there are two such
solutions with four $s=1/2$ 
singularities because the Hilbert space of four  ${s=1/2}$ spins has
two-dimensional zero-$\hat{\boldsymbol{S}}$ subspace:
$(1/2)^{\otimes 4} = 0\oplus 0 \oplus 1 \oplus 1\oplus 1 \oplus 2$,
where the spin representations are marked with their highest weights.

Besides, now we can explain what happens at $v_1 \to v_2$ when two
singularities of the solitons coalesce. In this limit the spin
operator (\ref{eq:T_operator}), 
$$
\hat{\boldsymbol{s}}(v)\to\frac{\hat{\boldsymbol{s}}_1 +
    \hat{\boldsymbol{s}}_2}{v-v_2}
  + \sum_{n\geq 3} \frac{\hat{\boldsymbol{s}}_n}{v-v_n}
  \qquad\qquad \mathrm{as}\quad
  v_1\,\to\,v_2\;,
$$
depends on the sum $\hat{\boldsymbol{s}}_1 +
\hat{\boldsymbol{s}}_2$. The corresponding solutions have
singularities at $v = v_2$ of powers $|s_1 - s_2|,\;  |s_1 -
s_2|+1,\;\dots,\; (s_1 + s_2)$ in accordance with the irreducible
representations of $\hat{\boldsymbol{s}}_1+\hat{\boldsymbol{s}}_2$.
For instance, consider coalescence of two $s_{1,2}=1/2$ singularities as
$v_1 \to v_2$. The second--order equations~(\ref{eq:31}) at 
these singularities have four solutions corresponding to four eigenstates of two
$s=1/2$ spins.  In the limit $v_1 \to v_2$ the spins  sum up and we
obtain\footnote{One can explicitly demonstrate    this  by solving
  Eqs.~(\ref{eq:31}) to the leading order in   ${v_1  - v_2 \to 0}$.} one
$s=0$ (non-singular) solution and  three solutions with 
$s=1$ singularity. 

Finally, one can obtain more general solutions with infinite number of
singularities using the thermodynamic Bethe Ansatz for the Gaudin
model~\cite{vanTongeren:2016hhc}. 

{\bf Example.} Consider the solution with four $s=1/2$ singularities
arranged in two complex conjugate pairs $v_{1,2} = a_1 \pm ib_1$,
$v_{3,4} = a_2 \pm ib_2$. Solving Eqs.~(\ref{eq:finiteness}),
(\ref{eq:add_constraint}), (\ref{eq:31}), we obtain, as expected
above, two solutions
\begin{multline}
  \label{eq:44}
  T_{vv}^{(\pm)}=\frac{6b_1^2\, 
    \mathrm{e}^{-2\phi_0}}{\left((v-a_1)^2+b_1^2\right)^2} 
  +\frac{6b_2^2 \, \mathrm{e}^{-2\phi_0}}{\left((v-a_2)^2+b_2^2\right)^2}
  \\-2\mathrm{e}^{-2\phi_0}\, \frac{(a_1-a_2)^2+b_1^2+b_2^2\pm\sqrt{\Delta}} 
  {\left((v-a_1)^2+b_1^2\right)\left((v-a_2)^2+b_2^2\right)}\;,
\end{multline}
where $\Delta=\left((a_1-a_2)^2+b_1^2+b_2^2\right)^2+12b_1^2
b_2^2>0$. In the limit $a_1 \to a_2$, $b_1 \to b_2$ the pairs of
singularities in the upper and lower parts of the complex $v$-plane coalesce,
and one obtains a nonsingular solution and a solution~(\ref{eq:40})
with two $s=1$ singularities,  
$$
T_{vv}^{(+)}\to 0 \;, \qquad\qquad 
T_{vv}^{(-)}\to \frac{16 \mathrm{e}^{-2\phi_0}\,
  b_2^2}{((v-a_2)^2+b_2^2)^2} \;,
$$
again in accordance with the above expectations.

\begin{figure}[htb]

  \vspace{-5.5mm}
  \centerline{\includegraphics[width=0.28\textwidth,angle=45]{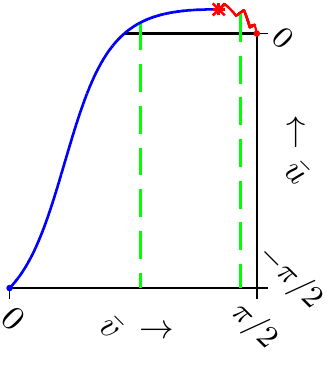}}

  \vspace{-4mm}
  \caption{Solution ``$-$'' in Eq.~(\ref{eq:44}) with four $s=1/2$
    poles and parameters $\lambda a_1 = -0.1$, $\lambda b_1 = 1$,
    $\lambda a_2 = 0.2$, $\lambda b_2 = 10$, and 
    $\mathrm{e}^{-2\phi_0}= 1$. The two peaks of the incoming matter flux
    are marked by the dashed lines. For this choice of parameters, the second
    peak forms the black hole.\label{fig:4pole_bh}} 
\end{figure}

Note that  $T_{vv}^{(+)}(v)$ is not positive-definite at real positive
$v$ and therefore unphysical. The function $T_{vv}^{(-)}(v)$ describes
incoming matter flux with two peaks at $v\sim a_1$ and $a_2$, see
Fig.~\ref{fig:4pole_bh}.

\subsection{Positivity condition}
\label{sec:positivity-condition}
Physical solutions have real $\psi(v)$ at real $v$. Thus, their
singularities $v_n$ and zeros $\tilde{v}_m$  are either real or
organized in complex  conjugate pairs
like in Fig.~\ref{fig:sing}. Besides, the singularities $v_n$ may not be
placed at the physical part $v\geq 0$ of the real axis.

The remaining nontrivial condition is $T_{vv}(v)\geq 0$ at $v\geq
0$, Eq.~(\ref{eq:37}). This inequality is not satisfied 
automatically. For example, our solutions with two
singularities~(\ref{eq:40}) have negative and positive $T_{vv}(v)$ 
at $v_{1,2} <0$ and $v_{1,2} = a\pm ib$,
respectively. In fact, any solution with all singularities placed 
at $v< 0$ is unphysical. In this case the operator
$\hat{\boldsymbol{s}}(v)$ at real $v$ is Hermitean, and therefore
$\hat{\cal T}(v)$ in Eq.~(\ref{eq:T_operator}) has 
positive-definite   eigenvalues~${{\cal T}(v) \propto
  -T_{vv}(v)}$.

In the opposite case when all singularities are organized in complex
conjugate pairs ${v_{2k-1},\, v_{2k} = a_{k} \pm ib_{k}}$ with
$s_{2k-1} = s_{2k}$, one 
expects to find at least one physical solution. Indeed, consider the
state $|\Psi_1\rangle$ (not an eigenstate) of the Gaudin model 
satisfying
${(\hat{\boldsymbol{s}}_{2k-1} + \hat{\boldsymbol{s}}_{2k}) 
| \Psi_1\rangle = 0}$ for all $k$. Explicit calculation shows that $\langle \Psi_1
| \hat{\cal T}(v) |\Psi_1 \rangle <0$ at real $v$. On the other hand, the
variational principle implies that for any $N$ real points $w_n$  there
exists  an eigenstate  $|\Psi\rangle$ minimizing all $\langle \Psi | 
\hat{\cal T}(w_n)|\Psi \rangle$. The respective eigenvalue ${\cal
  T}(v)$ is negative at all $v = w_n$ suggesting that ${T_{vv}(v) \propto
  -{\cal T}(v)}$ is positive at the entire real axis. 

Let us explicitly select the above physical solution at $b_{k} \to
0$. In this case $T_{vv}(v)$ falls into a collection of peaks at
$v\sim a_k$ near the singularities 
$v_{2k-1}$, $v_{2k}$. At $|v - a_k| \gg b_k$ and yet, far away from  
other singularities, the operator (\ref{eq:T_operator}) takes the
form $\hat{\cal T}(v) \approx (\hat{\boldsymbol{s}}_{2k-1} +
\hat{\boldsymbol{s}}_{2k})^2/(v-a_{k})^2$. Its eigenvalue ${\cal 
  T}(v) \propto -T_{vv}(v)$ is positive-definite unless the eigenstate
satisfies $(\hat{\boldsymbol{s}}_{2k-1} + \hat{\boldsymbol{s}}_{2k})
|\Psi\rangle = 0$. Thus, in the limit $b_k \to 0$ the physical
eigenstate coincides with the state $|\Psi_1\rangle$ introduced
above. The respective energy flux $T_{vv}(v)$ is the sum of
two-spin terms~(\ref{eq:40}),  
$$
T_{vv} \approx 8\mathrm{e}^{-2\phi_0} \sum_{k=1}^{N/2}
    \frac{s_{2k}(s_{2k}+1) b_k^2}{\left[(v  - a_k)^2 +
        b_k^2\right]^2}\qquad \mbox{at small $b_k$}\;.
$$
    One expects that this solution remains physical at finite $b_k$.

\begin{figure}[htb]
  
  \vspace{7mm}
  \hspace{2.25cm} (a) \hspace{7.5cm}(b)

  \vspace{-7mm}
  \centerline{\begin{minipage}{.28\textwidth}
      \includegraphics[width=\textwidth]{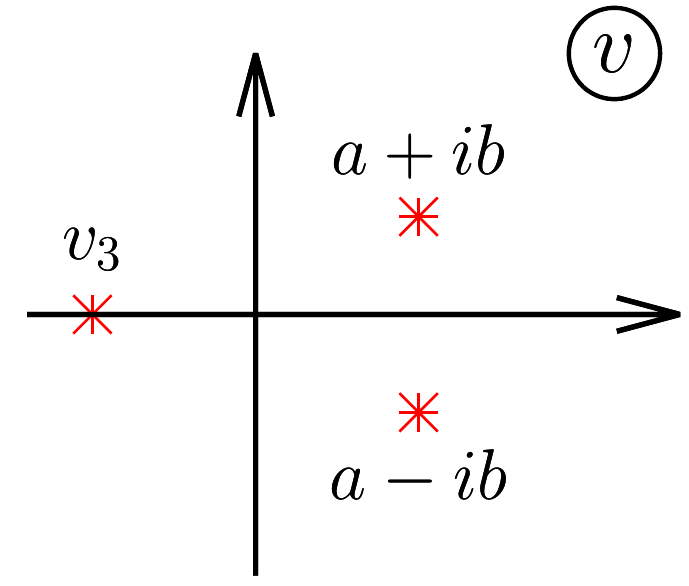}
    \end{minipage}\hspace{1cm}
    \begin{minipage}{.5\textwidth}
      \includegraphics[width=\textwidth]{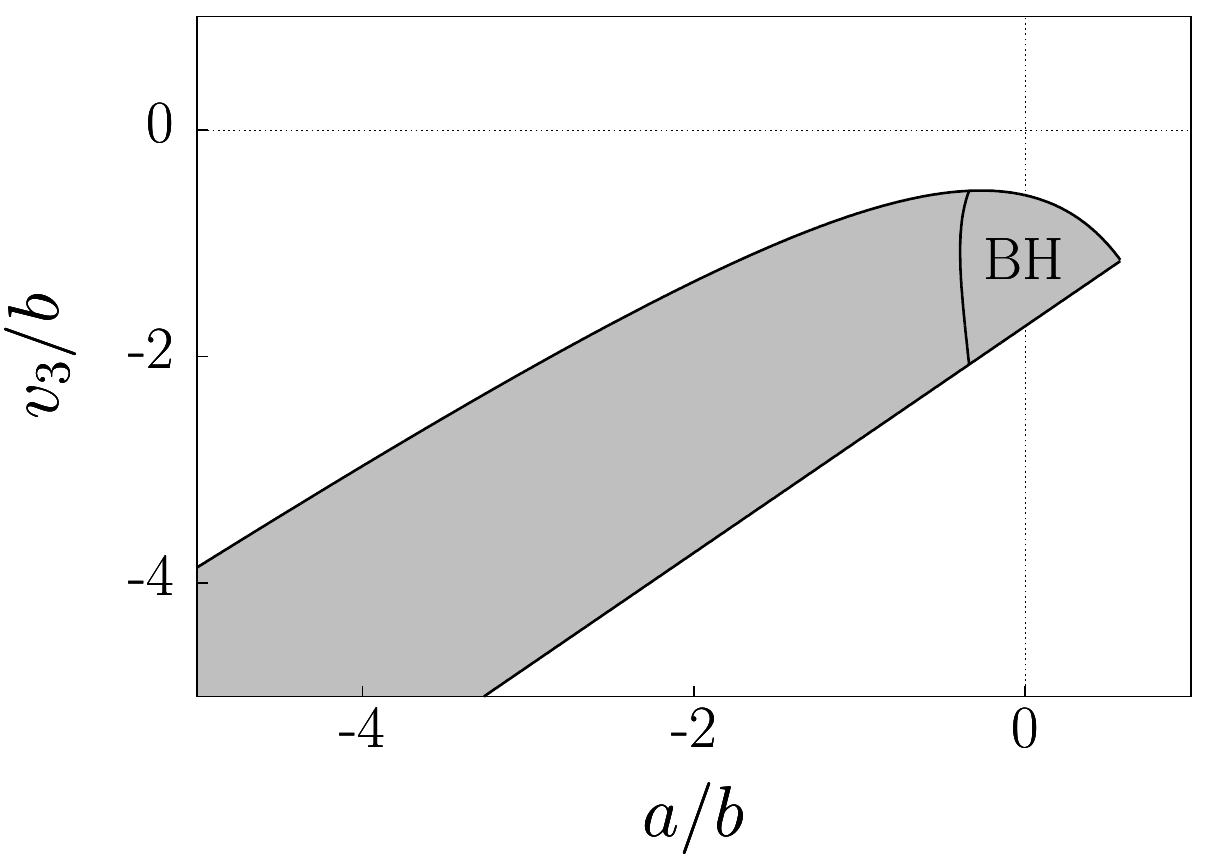}
  \end{minipage}}

  \caption{(a) Singularities of the solution (\ref{eq:45}). (b)
    Parameters of this solution giving positive-definite
    $T_{vv}(v)$ at $v\geq 0$ (gray region). The upper right corner of
    this region corresponds to black hole
    formation. \label{fig:3pole}}
\end{figure}

{\bf Example.} In general case the positivity condition bounds 
parameters of the solutions. Consider e.g.\ the soliton with three
$s=1$ singularities at $v_{1,2} = a\pm ib$, $v_3 <0$, see
Fig.~\ref{fig:3pole}a. Solving Eqs.~(\ref{eq:finiteness}),
(\ref{eq:add_constraint}), one obtains,  
\begin{equation}
  \label{eq:45}
  T_{vv}=\frac{16 \mathrm{e}^{-2\phi_0}b^2}{((v-a)^2+b^2)^2}
  -\frac{4\mathrm{e}^{-2\phi_0}\left[(a-v_3)^2+b^2\right]}{(v-v_3)^2
    \left[(v-a)^2+b^2\right]}\;.
\end{equation}
The second (negative) term in this expression represents contribution
of the singularity $v_3<0$. It can be compensated by the first term if
the singularities $v_{1}$ and $v_2$ are close enough to $v_3$. Namely, the
function (\ref{eq:45}) is positive-definite at $v\geq 0$ if
$a-b\sqrt{3} \leq v_3 \leq (a^2 + b^2)/(a-b\sqrt{3})$, see the gray region  
in  Fig.~\ref{fig:3pole}b. The solutions with these
parameters involve one peak of the incoming flux, just like the
solutions~(\ref{eq:40}). 

\section{Critical chaos}\label{sec:instability}
\subsection{Perturbative expansion in the critical regime}
\label{sec:pert-expans-crit}
In Sec.~\ref{sec:simpl-bound-equat} we argued that the critical
solutions at the verge of black hole formation have
constant $\psi(v)$ and null boundary $U(v)$ at large $v$,
see Fig.~\ref{fig:psi}. One can say that they describe formation of
the minimal-mass black holes with the boundary placed precisely at the
horizon~\cite{Kiem}\cite{Peleg:1996ce}, cf.~\cite{Choptuik:1992jv,
  Gundlach:2002sx}.

\begin{figure}[htb]
  \centerline{\includegraphics[width=0.7\textwidth]{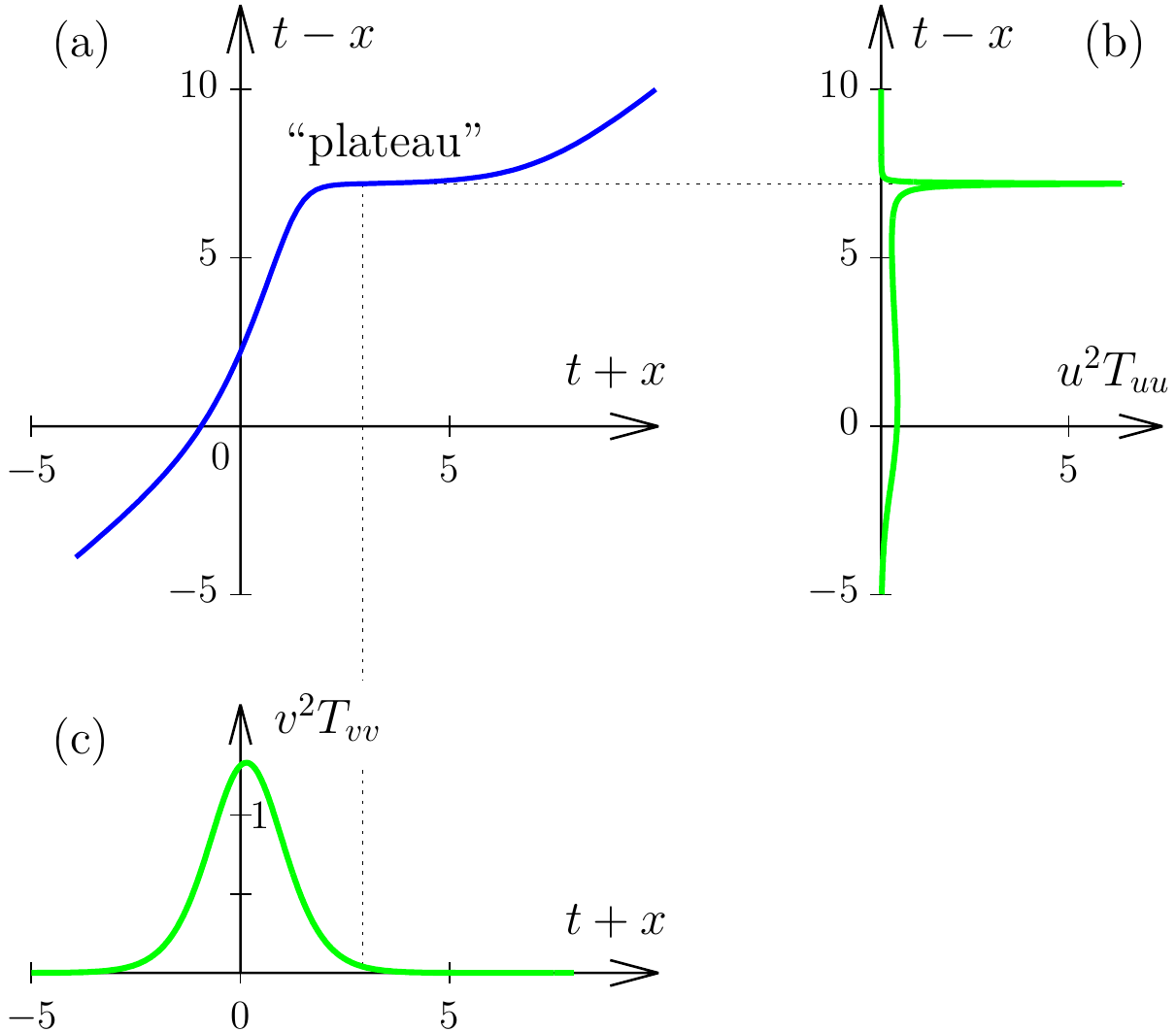}}
  \caption{Solution (\ref{eq:40}) at almost critical values of   
    parameters $s=b = \mathrm{e}^{-2\phi_0} = 1$ and $a =
    a_{cr} - 10^{-3}$,   where $a_{cr} =- 1/\sqrt{3}$ and we use
    units with $\lambda = 1$. In this case $C
    \approx 7 \times 10^{-4} \ll 1$. Figure (a) shows the boundary $u
    = U(v)$ in the asymptotically flat light-cone coordinates
     $t+x = \log(\lambda v)/\lambda$, $t-x =
    -\log(-\lambda u)/\lambda$, see Eq.~(\ref{eq:15}). In Figs.~(b),
    (c) we plot the outgoing 
    and incoming energy fluxes $u^2 T_{uu}$ and $v^2 T_{vv}$ as
    functions of $t-x$ and $t+x$, respectively. \label{fig:thunderpop}}
\end{figure}

At energies somewhat below critical the boundary has long
almost null part (``plateau''), see Fig.~\ref{fig:thunderpop}a. The
energy flux reflected from this part is strongly amplified by the
Lorentz factor of the boundary and forms a high and narrow peak in
$T_{uu}(u)$, see Fig.~\ref{fig:thunderpop}b. We will argue below 
that in the critical limit the peak tends to a $\delta$-function 
(shock-wave)  with energy equal to the minimal black hole mass
$M_{cr}$. In the overcritical solutions the shock-wave is
swallowed by the black hole. Besides, we will see in the next Section
that the structure of the peak is highly sensitive to
the initial data. This feature impedes global integrability of the
model. 

Let us find the boundary $U(v)$ in the ``plateau'' region where $v$ is
large and $T_{vv}(v)$ is small. In this case Eq.~(\ref{eq:schrodi})
can be solved perturbatively by representing  $\psi = 1+\psi^{(1)} +
\psi^{(2)} + \dots$, where $\psi^{(k)} \propto (T_{vv})^k$. 
Using $\psi \approx 1$ in the r.h.s.\ of Eq.~(\ref{eq:schrodi}), we obtain, 
\begin{equation}
  \label{eq:pert_psi1}
  \psi^{(1)}(v) =  Cv + e^{2\phi_0}\left[g(v)-g_\infty\right]\;, 
\end{equation}
where the function $g(v)$ is introduced in Eq.~(\ref{eq:g_h_def}) and
$g_{\infty}$ is its value at $v\to +\infty$. Note that the linear
asymptotics $Cv \ll 1$ of the solution appears at first order of
expansion  in Eq.~(\ref{eq:pert_psi1}) because in the near-critical
regime $\partial_v \psi  
\approx  C$ is small at large $v$. In what follows we will regard $C$
as a parameter of the expansion. Using $\psi \approx 1+\psi^{(1)}$
in the r.h.s.\ of Eq.~(\ref{eq:schrodi}), we get
$$
\partial_v \psi^{(2)}(v)
  =\mathrm{e}^{2\phi_0}(g-g_\infty)
  \left(\mathrm{e}^{2\phi_0} \partial_v g - C\right) +
  \mathrm{e}^{2\phi_0}C\, v\partial_v g +
  e^{4\phi_0}\int\limits^{\infty}_v\;dv'\left(\partial_{v'} 
g\right)^2\;. 
$$
The higher-order corrections $\psi^{(n)}$ are obtained in similar
way.

Now, we compute the reflected energy
flux $T_{uu}(u)$ and the boundary function $U(v)$ using
Eqs.~\eqref{eq:16} and \eqref{eq:13}, 
\begin{align}
  &T_{uu}(U(v)) \approx \frac{ \lambda^4 \mathrm{e}^{4\phi_0}\, T_{vv}(v)
  }{\left[C+e^{2\phi_0}\partial_v g(v)\right]^4}\;,\label{eq:final_pulse}\\
  & \lambda^2 U(v) \approx  -\mathrm{e}^{-2\phi_0} C +
    \mathrm{e}^{-2\phi_0} C^2 v   + 2 C(g - g_\infty)
  - \mathrm{e}^{2\phi_0} \int_v^{\infty} dv' \,
  \left(\partial_{v'} g\right)^2\;.
     \label{eq:46} 
\end{align}
We kept one and two orders of the expansion in
Eqs.~\eqref{eq:final_pulse} and~\eqref{eq:46}, respectively. Note that
the leading (first) term in 
$U(v)$ is constant; this behavior  corresponds to the ``plateau'' in
Fig.~\ref{fig:thunderpop}a. At the same time, the reflected flux
(\ref{eq:final_pulse}) has a peak at large $v$ corresponding to
$\partial_v g \sim C\mathrm{e}^{-2\phi_0}$. This peak is narrow in
terms of slowly-changing $u = U(v)$ in agreement with
Fig.~\ref{fig:thunderpop}b. 

Using the soliton asymptotics
$T_{vv} \propto v^{-4}$ and $\partial_v g \propto v^{-3}$, one finds
that the peak in Eq.~\eqref{eq:final_pulse} occurs at $v \propto
C^{-1/3}$, and its width $\Delta v$ is of the same order. The respective value of
$U(v)$ is approximately given by the first term in Eq.~(\ref{eq:46}),
while the peak width $\Delta U \propto C^{2/3} U$ is
controlled by the second-order terms. In  the critical limit $C\to 0$
the peak of  $T_{uu}(u)$ is infinitely high and narrow. 

Calculating the total energy within the shock-wave  at 
$C\to 0$, we obtain,  
$$
E_{\mathrm{peak}} = \lambda \int\limits_{u\sim C} |u| du \,
  T_{uu}(u) \to -2\lambda C\int^{+\infty}_0
  \frac{ dv\, \partial_v^2 g(v)}{\left[C+e^{2\phi_0}\partial_v g(v)\right]^2}
 \to  2 \lambda \mathrm{e}^{-2\phi_0} 
$$
where Eqs.~(\ref{eq:final_pulse}), (\ref{eq:46}) were used. The value
of $E_{\mathrm{peak}}$ coincides with the minimal black hole mass
$M_{cr}$ implying 
that the peak of $T_{uu}(u)$ tends to a
$\delta$-function in the critical limit.

\subsection{Shock-wave instability}
\label{sec:thund-inst}
Since our model is equipped with the general solution,  one may think
that it is integrable, i.e.\ has a complete set of conserved
quantities $\{ I_k\}$ smoothly foliating the phase space. In the
in-sector these quantities are arbitrary functionals $I_k[f_{in}]$ of
conserved $f_{in}(v)$, cf.~\cite{Bazhanov:1994ft}. Then,  $I_k$ can be computed 
at arbitrary space-like line: to this end one evolves the classical
fields from this line to $J^-$, extracts the incoming wave\footnote{Recall   that all
  our solutions start from flat space-time in the
  infinite   past.} $f_{in}(v)$, and calculates $I_k[f_{in}]$. The
quantities $\{I_k\}$ obtained in this way are conserved by definition. For
example, in the out-sector one gets $I_k[f_{out}] \equiv I_k[f_{in}]$ if
$f_{out}(u)$ and $f_{in}(v)$ are related by classical evolution,
Eq.~\eqref{eq:riccati}.   

Let us argue, however, that  $\{I_k\}$ cannot be smoothly
defined in the near-critical regime because the map
$f_{in}\to f_{out}$ in this case is essentially singular.
To simplify the argument, we consider solutions with the
modulated flux  at large~$v$,
\begin{equation}
  \label{eq:49}
  T_{vv}  = (\partial_v f_{in})^2\;,
  \qquad \partial_v f_{in}\approx A\, v^{-2}\cos(\omega\ln\,(\lambda
  v)) \qquad \mbox{at} 
\qquad v \gtrsim C^{-1/3}\;,
\end{equation}
where $C$ is the small parameter of the near-critical expansion. If
$\omega$ is small as 
well, the asymptotics of $T_{vv}$ is almost power-law, like in the 
ordinary solitons. However, the shock-wave
part of the reflected flux represents squeezed and amplified tail of
$T_{vv}$ at $v \sim C^{-1/3}$, 
see Fig.~\ref{fig:thunderpop}. It should be essentially modulated. For
simplicity, let us   characterize the outgoing wave packet with a
single quantity
\begin{align}
  \notag
&{\cal I}_3(C,\, \omega)  \equiv \int\limits_{-\infty} ^{+\infty}
        d(t-x) \left(\partial_{t-x} f_{out}\right)^3 = 
        \Delta {\cal I}_3(C,\, \omega) + \mbox{const}\;,\\
          \label{eq:48}
       & \Delta {\cal I}_3 =
        \int\limits_{0}^{\infty} dv\;\frac{C^2
          (\partial_v f_{in})^{3}}{\left[C+e^{2\phi_0}g_v(v)\right]^4}\;, 
\end{align}
where we used the flat coordinates~\eqref{eq:15} in the
definition of ${\cal I}_3$, then separated the shock-wave part 
$\Delta {\cal I}_3$ of the integral at $t - x \equiv -\log(-\lambda 
u)/\lambda\gtrsim \log C$ from the $(C,\, \omega)$-independent
contribution at smaller $t-x$. In the second line we
substituted the  shock-wave 
profile~\eqref{eq:final_pulse}, \eqref{eq:46} and extended the
integration range to $v\geq 0$. Now, one substitutes the
asymptotics~(\ref{eq:49}) into Eq.~(\ref{eq:48}) and finds that 
$\Delta {\cal I}_3(C,\, \omega)$  is quasi-periodic. Indeed,
change of the integration variable $v\,\mapsto\,ve^{2\pi n/\omega}$
with integer $n$ gives relation\footnote{In this case 
  $g'(v\mathrm{e}^{2\pi   n/\omega}) = \mathrm{e}^{-6\pi n/\omega}
  g'(v)$, where the derivative is taken with respect to the argument, see 
  Eqs.~(\ref{eq:g_h_def}).} 
$\Delta {\cal I}_3(e^{6\pi n/\omega}C,\, \omega)=e^{-2\pi n/\omega}\Delta
{\cal I}_3(C,\, \omega)$. Thus,
$\Delta {\cal I}_3 = C^{-1/3} \, {\cal J}(\omega \log C)$,  where
${\cal J}(x)$ is $6\pi$-periodic.  

\begin{figure}[htb]
\center{\includegraphics[width=0.45\linewidth]{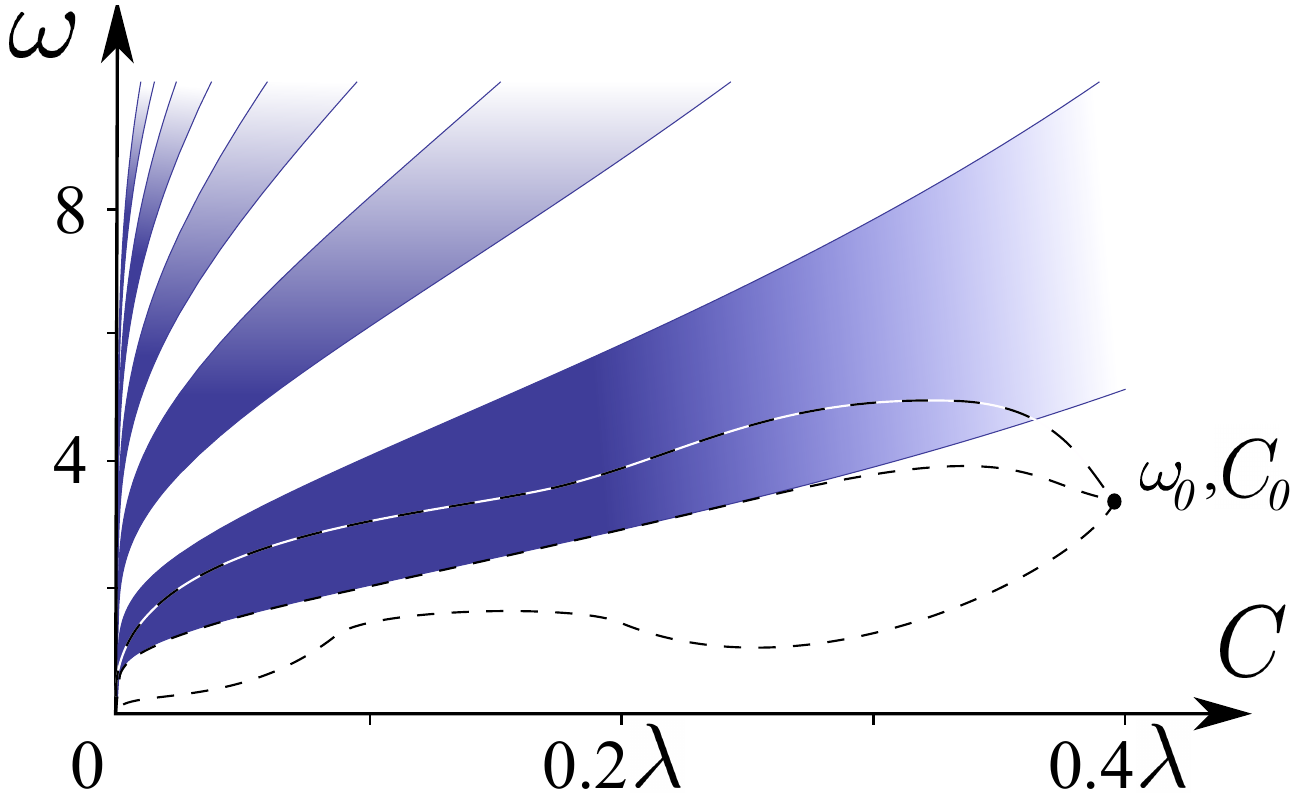}} 
\caption{Regions $\Delta {\cal I}_3 >0$ (white) and $\Delta {\cal
    I}_{3}<0$ (blue) in the $(C,\,\omega)$ plane. We use the solution 
  (\ref{eq:49}) with $A^2 =
  12\mathrm{e}^{-2\phi_0}/\lambda^2$. \label{fig:nonint} } 
\end{figure}

We see that $\Delta {\cal I}_3$  has an essential singularity at
$\omega =  C = 0$. Indeed, taking the limit $C\to 0$ along the paths
$\omega \log C = \mbox{const}$, one obtains $\Delta {\cal I}_3 \to
-\infty,\, 0$, or $+\infty$, see Fig.~\ref{fig:nonint}. Thus,
any value of $\Delta {\cal I}_3$ can be obtained by adjusting the
limiting path.

The above property ascertains dynamical chaos in the critical limit
of our model. Indeed, infinitesimally small changes~\eqref{eq:49} of
the initial data at small $C$ produce outgoing fluxes with
essentially different  values of ${\cal I}_3$. This prevents one from
characterizing the critical evolution with a  set of smooth
conserved quantities $I_k$. Indeed, all functionals $I_k[f_{in}]$,
being smooth in the in-sector, are  not sensitive to $\omega$ at small
values of  latter. Thus, they fail to describe essentially different
out-states $f_{out}(u)$ at different $\omega$. From a more general
perspective, one can introduce the integrals which are smooth either
in the in-sector or in the out-sector, but not in both.

\section{Discussion}\label{sec:conc}
In this paper we considered two-dimensional CGHS model with a
regulating dynamical boundary~\cite{Chung:1993rf,
  Schoutens:1994st}. This model is weakly coupled and causally similar
to the spherically-symmetric gravity in many dimensions. We
demonstrated that classical field equations  in this model are
exactly solvable. We constructed their general solution 
and studied in detail a large subset of soliton solutions
with transparent properties. We illustrated  the results with many
explicit examples hoping that this model will serve as a practical
playground for black hole physics.  

In the critical regime i.e.\ at the verge of black hole formation, our
model displays dynamical instabilities  specific to chaotic 
systems. This property is similar to the near-horizon chaos
suggested recently in the context of AdS/CFT
correspondence~\cite{Shenker:2013pqa, Shenker:2014cwa,
  Polchinski:2015cea, Hashimoto:2016dfz, Maldacena:2015waa, 
  Turiaci:2016cvo}. We argued that it hinders global
integrability of the model.

We see several applications of our results.
First, exact solvability may extend to one-loop semiclassical level
if one adds a reflective boundary to the RST
model~\cite{Russo:1992ax}. This approach,  if successful, will produce
analytic solutions describing black hole formation and evaporation. The
singularities of such solutions should be either covered by the 
boundary or hidden behind the space-like line $\phi =\phi_0$, see
Fig.~\ref{fig:psi}b. Then a complete Penrose diagram for the
evaporation process may be obtained,
cf.~\cite{Das:1994kr,Bose:1995bk,Bose:1996pi,Peleg:1996ce}.

Second, in the alternative approach one directly adds one-loop
corrections to the classical equations of our model with a boundary and  integrates
the resulting system numerically, cf.~\cite{Ashtekar:2010hx,
  Ashtekar:2010qz}. By the same reasons as above, the respective
solutions should completely describe the process of black hole
evaporation.

Third and finally, the model of this paper is ideal for applying the 
semiclassical method of~\cite{Levkov:2007yn, Bezrukov:2015ufa} which relates
calculation of the exponentially suppressed $S$-matrix elements
to certain complex classical solutions. The results of such
calculations may be used to test unitarity of the gravitational 
$S$-matrix~\cite{Bezrukov:2015ufa}.   

\paragraph{Acknowledgments.}
We thank S.~Sibiryakov for participating at early
stages of this project. We are grateful to \'Ecole Polytechnique
F\'ed\'erale de Lausanne for hospitality during our visits. This
work was supported by the grant RSCF 14-22-00161. 

\appendix
\section{Field equations and boundary conditions}
\label{AppGH}
\subsection{Derivation}
\label{sec:field-equat-bound}
Field equations in the bulk are obtained by varying the action
(\ref{eq:cghsmirror_action}) with respect to $g_{\mu\nu}$,
$\phi$, and $f$, and ignoring the boundary terms,  
\begin{align}
  \label{eq:1}
  4e^{-2\phi}\nabla_\mu\nabla_\nu\phi + 4g_{\mu\nu} e^{-2\phi}
  \left[(\nabla\phi)^2 - \nabla^2\phi-\lambda^2\right]
  & = \nabla_\mu f \nabla_\nu f - \frac12 g_{\mu\nu} (\nabla
  f)^2\;,\\
  \label{eq:2}
(\nabla\phi)^2- \nabla^2\phi-\lambda^2& =R/4\;,\\
  \nabla^2 f& =0\;. \label{eq:3}
\end{align}
The first line here relates the energy-momentum tensors of
$\phi$ and $f$, $-T_{\mu\nu}^{(\phi)} = T_{\mu\nu}^{(f)}$. The second
line implies, in addition, that the rescaled metric
$\mathrm{e}^{-2\phi}g_{\mu\nu}$ is flat.

To find the boundary conditions at the line $\phi = \phi_0$, we keep
the boundary terms in the variation of the action. For a start, let us 
consider variations preserving the coordinate position of the
boundary $\phi = \phi_0$. We  take  $\delta \phi = 0$ along this line and fix the
direction of its outer normal, $\delta n_\mu \propto n_\mu$. The  
integration  domains in Eq.~(\ref{eq:cghsmirror_action}) are unchanged 
by such variations. One obtains,
\begin{equation}
  \label{eq:4}
  \delta S = - \int\limits_{\phi=\phi_0} d\tau \, \left[ 2  h^{\mu\nu} \delta
  h_{\mu\nu} \, \mathrm{e}^{-2\phi_0}\left(n^{\kappa} \nabla_\kappa
  \phi - \lambda \right) +  \delta f \, n^\kappa \nabla_\kappa f
  \right] = 0\;,
\end{equation}
where we canceled the bulk terms using
Eqs.~(\ref{eq:1})---~(\ref{eq:3}) and introduced the induced metric
$h_{\mu\nu} \equiv g_{\mu\nu} - n_\mu n_\nu$. The variation
(\ref{eq:4}) gives the boundary conditions~\eqref{eq:neumann}. Note
that the space-time is flat near the
boundary: one obtains $R=0$ at $\phi  = \phi_0$ using the first of
Eqs.~\eqref{eq:neumann}, Eq.~(\ref{eq:2}) and the trace of
Eq.~(\ref{eq:1}).  

Now, let us consider general variations shifting the position of the
boundary. They are combinations of the general coordinate
transformations and position-pre\-ser\-ving variations considered
above. The action is unchanged by these variations: it is
covariant and already extremized at fixed coordinate position of the boundary. 

\subsection{Solution in the conformal gauge}
\label{sec:quasi-kruskal-gauge}
Let us review the general solution~\cite{Callan:1992rs} of the bulk
equations~(\ref{eq:1})---~(\ref{eq:3}), see ~\cite{Giddings:1994pj,
  Strominger:1994tn} for details.

In the light-cone frame (\ref{eq:6}) Eq.~(\ref{eq:3}) takes the
form  $\partial_u \partial_v f = 0$, its solution is given by
Eq.~(\ref{eq:bulk_scalar}). Combining Eq.~(\ref{eq:2}) with the trace
of Eq.~(\ref{eq:1}) and substituting $R = 8  \mathrm{e}^{-2\rho}
\partial_u \partial_v \rho$, we obtain, 
$$
  \partial_u \partial_v (\phi - \rho) = 0 \qquad \Rightarrow \qquad
  \phi = \rho\;,
$$
where the residual coordinate freedom\footnote{Namely, the 
  transformations $u\to \tilde{u}(u)$, $v\to \tilde{v}(v)$ preserving
  the metric (\ref{eq:6}).} was fixed in the last equation. After
that Eqs.~(\ref{eq:1}), namely, 
\begin{align}
  \notag
   \partial_u^2 \mathrm{e}^{-2\phi} &= - \frac{1}{2}\, (\partial_u
   f)^2\;,\\
   \notag
   \partial_v^2 \mathrm{e}^{-2\phi} &= - \frac{1}{2}\, (\partial_v f)^2\;,\\
   \notag
   \partial_u \partial_v \mathrm{e}^{-2\phi} &= -\lambda^2\;,
\end{align}
are integrated into
\begin{equation}
  \label{eq:9}
\mathrm{e}^{-2\rho} =   \mathrm{e}^{-2\phi} = \frac{M_-}{2\lambda}-
\lambda^2 (u-u_0) (v-v_0) + g(v) + h(u) \;.
\end{equation}
In this expression $M_-$, $u_0$, and $v_0$ are integration
constants; functions $g(v)$ and $h(u)$ were introduced in 
Eq.~(\ref{eq:g_h_def}). We fix $u_0 = 
v_0 = 0$ by shifting $u$ and $v$. After that $M_-$ represents the
mass of white hole in the infinite past~\cite{Giddings:1994pj,
  Strominger:1994tn}. Indeed, 
the past time infinity $i^-$ in Fig.~\ref{fig:penrose-intro}b is
reached at $u\to -\infty$, $v\to 0$, 
and constant $\phi$, cf.\ Eq.~(\ref{eq:10}). If $M_-\ne 0$, the curvature
remains nonzero in this limit,
$$
R =  4 \mathrm{e}^{2\rho}(\partial_u \mathrm{e}^{-2\rho})
    (\partial_v \mathrm{e}^{-2\rho}) - 4
    \partial_u \partial_v \mathrm{e}^{-2\rho}  \to 2\lambda
  \mathrm{e}^{2\phi} M_-\;,
$$
where Eq.~(\ref{eq:9}) with $u_0 = v_0 = 0$ was used. In this paper we
consider solutions starting from flat space-time. Thus, $M_- =0$,
and Eq.~(\ref{eq:9}) takes the form 
(\ref{eq:bulk_gravity}).

It is worth noting that the patch $u \in (-\infty,\, 0)$ and $v \in
(0,\, +\infty)$ covers all space-time accessible to the outside
observer. Indeed, we already mentioned that the time infinities
$i^-$ and $i^+$ are reached in the limits $u \to -\infty$ and
$v\to +\infty$ at finite values of the dilaton field $\phi$. By
Eq.~(\ref{eq:bulk_gravity}), the product $uv$ remains finite in these
limits implying $v\to +0$ as $u \to -\infty$ ($i^-$) and $u \to -0$ as $v\to
+\infty$ ($i^+$), see Fig.~\ref{fig:coordinates}.

We proceed by deriving equation of motion for the boundary $u =
U(v)$ satisfying $\phi(U(v),\, v) = \phi_0$. Taking the derivative of
Eq.~(\ref{eq:bulk_gravity}) along this line, we find, 
\begin{equation}
  \label{eq:12}
0 =  \frac{d}{dv}\, \mathrm{e}^{-2\phi_0}= U'\,  \left[ \partial_u h - \lambda^2
v\right]+ \partial_v g  - \lambda^2 U\;, \qquad\mbox{at} \qquad u=U(v)\;,
\end{equation}
where $U' \equiv dU/dv >0$ because the boundary is time-like. The other
two equations come from the boundary
conditions~(\ref{eq:neumann}). Introducing the unit outer normal   
$$
n^u = \mathrm{e}^{-\phi_0} \sqrt{U'}\;, \qquad\qquad 
n^v = - \mathrm{e}^{-\phi_0} / \sqrt{U'}
$$
and using Eq.~(\ref{eq:12}), we rewrite Eqs.~(\ref{eq:neumann}) in the
form~(\ref{eq:riccati}).  

At this point, we have three equations, Eqs.~\eqref{eq:12} and
(\ref{eq:riccati}), for the two unknown functions 
$f_{out}(u)$ and $U(v)$. Note, however, that Eq.~\eqref{eq:12} follows
from the other two equations. Indeed, 
$$
\frac{d}{dv} \left(\frac{\partial_v g - \lambda^2 U}{U'}\right) =
  \lambda^2 \mathrm{e}^{-2\varphi_0}\frac{d}{dv}(\partial_v g - \lambda^2 U)^{-1}
   = \frac{(\partial_v f_{in})^2/2 + \lambda^2 U'}{U' } = 
  \frac{d}{dv} (\lambda^2  v-\partial_u h)\;,
$$
where we expressed $U'$ and $g$ via Eqs.~(\ref{eq:riccati}) and
(\ref{eq:g_h_def}) in the first and second  equalities, then turned
$f_{in} \to f_{out}$ by the second of Eqs.~(\ref{eq:riccati}) and used
the equation for $U'$, again. One 
concludes that Eq.~\eqref{eq:12} is automatically satisfied once the
initial conditions for~$U(v)$ are chosen correctly.

\section{Bethe Ansatz for the Gaudin model}
\label{sec:gaudin-spin-chain}
In this Appendix we review Bethe Ansatz for the Gaudin
model~(\ref{eq:gaudin_hamiltonian}),
see~\cite{Gaudin:1976sv, Feigin:1994in, Frenkel:1995zp} for
details.

One introduces raising and lowering operators  $\hat{s}^\pm (v) =
\hat{s}^1 (v) \pm i \hat{s}^2(v)$ for the position-dependent spin
(\ref{eq:T_operator}). The commutation rules of these operators
are 
$$
[\hat{s}^-(v),\,\hat{s}^+(w)]=2\,\frac{\hat{s}^3(v)-\hat{s}^3(w)}{v-w}\;,
\qquad
    [\hat{s}^3(v),\,\hat{s}^{\pm}(w)] =
    \mp\frac{\hat{s}^\pm(v)-\hat{s}^\pm(w)}{v-w}\;.
$$
The Hamiltonian $\hat{\cal T}(v)$ in Eq.~(\ref{eq:T_operator}) takes
the form 
\begin{equation}\label{eq:T_operator_s}
 \hat{{\cal T}}(v)=\frac12\hat{s}^+(v)\hat{s}^-(v) +
\frac12\hat{s}^-(v)\hat{s}^+(v) + (\hat{s}^3(v))^2\;.
\end{equation}
Now, it is straightforward to check that the spin-down
state~(\ref{eq:gaudin_vacuum}) is an eigenstate:
$$
  \hat{\cal T}(v) |0\rangle = \left[(W_0)^2 + \partial_v W_0\right]
  |0\rangle\;,
  \qquad \mbox{where} \qquad W_0(v) = -\sum_n \frac{s_n}{v - v_n} 
$$
is the eigenvalue of the third spin component,
$\hat{s}^{3}(v) | 0\rangle = W_0(v)|0\rangle$. 

One explicitly acts with $\hat{\cal T}(v)$,
Eq.~(\ref{eq:T_operator_s}), on the state~(\ref{eq:bethe_states})
 and obtains,
\begin{equation}
  \label{eq:5}
\hat{\cal T}(v)|\tilde{v}_1,\dots,\tilde{v}_M\rangle=
   {\cal T}(v)|\tilde{v}_1,\dots,\tilde{v}_M\rangle - \sum_{m=1}^M 
    \frac{2L_m}{v - \tilde{v}_m}|\tilde{v}_1,\dots, \tilde{v}_m \,
    \mapsto\,v,\dots , \tilde{v}_M\rangle\;,
\end{equation}
where ${\cal T}(v)$ is given by Eq.~(\ref{eq:39}), $L_m$ is the
left-hand side of Eq.~(\ref{eq:41}), and arrow denotes
substitution. Note that the relations 
\begin{align}
& [\hat{{\cal T}}(v),\,\hat{s}^+(w)] = \frac{2}{v-w} \left(
  \hat{s}^+(w) \hat{s}^3(v) - \hat{s}^+(v) \hat{s}^3(w) \right)
  \;,\notag\\
&
  \hat{s}^3(v)| \tilde{v}_1, \dots, \tilde{v}_M \rangle = W(v) 
  |\tilde{v}_1, \dots, \tilde{v}_M \rangle - \sum_m
  \frac1{v-\tilde{v}_m} |\tilde{v}_1, \dots, \tilde{v}_m\, \to\, v,
  \dots, \tilde{v}_M \rangle\notag \;,
\end{align}
where $W(v)$ is defined in Eq.~(\ref{eq:39}), are helpful for deriving
Eq.~\eqref{eq:5}.

We conclude that Eq.~(\ref{eq:5}) coincides with the eigenproblem for
$\hat{\cal T}(v)$ if the Bethe equations $L_m= 0$, Eqs.~(\ref{eq:41}),
are satisfied.  In this case the Bethe states~(\ref{eq:bethe_states}) are 
the eigenstates of the Gaudin
Hamiltonians~(\ref{eq:gaudin_hamiltonian}). Moreover, one can
prove~\cite{Gaudin:1976sv, Feigin:1994in, Frenkel:1995zp} 
that the basis~(\ref{eq:bethe_states}) is complete. 


\end{document}